\documentclass[showpacs,amsmath,amssymb,aps,showkeys,floatfix,prd,a4paper]{revtex4} 
\usepackage[utf8]{inputenc} %
\usepackage{verbatim} 
\usepackage[dvips]{graphicx}
\usepackage{dcolumn}
\usepackage{bm}
\usepackage{epsfig}
\usepackage{amsfonts}
\usepackage{amssymb,amscd}

\newcommand{\be}{\begin{equation}}
\newcommand{\ee}{\end{equation}}
\newcommand{\ben}{\begin{eqnarray}}
\newcommand{\een}{\end{eqnarray}}

\begin{document}

\title{$J/\psi$ regeneration in a hadron gas: an update}
\author{ L. M. Abreu\footnote{luciano.abreu@ufba.br}}   
\affiliation{Instituto de F\'isica, Universidade Federal da Bahia, 
Campus Universit\'ario de Ondina, 40170-115, Bahia, Brazil}
\author{K.~P.~Khemchandani\footnote{kanchan@if.usp.br}}
 \affiliation{
Universidade Federal de S\~ao Paulo, C.P. 01302-907, S\~ao Paulo, Brazil
}
\author{A.~Mart\'inez~Torres\footnote{amartine@if.usp.br}}
\affiliation{
Instituto de F\'isica, Universidade de S\~ao Paulo, C.P. 66318, 05389-970 S\~ao 
Paulo, S\~ao Paulo, Brazil
}
\author{ F.~S.~Navarra\footnote{navarra@if.usp.br}}
 \affiliation{
Instituto de F\'isica, Universidade de S\~ao Paulo, C.P. 66318, 05389-970 S\~ao 
Paulo, S\~ao Paulo, Brazil\\
Institut de Physique Th\'eorique, Universit\'e Paris Saclay,\\
CEA, CNRS, F-91191, Gif-sur-Yvette, France
}
\author{ M.~Nielsen\footnote{mnielsen@if.usp.br} }
 \affiliation{
Instituto de F\'isica, Universidade de S\~ao Paulo, C.P. 66318, 05389-970 S\~ao 
Paulo, S\~ao Paulo, Brazil\\
SLAC Nacional Acelerator Laboratory, Stanford University, Stanford, 
California 94309, USA}

\begin{abstract}

In heavy ion collisions after the quark-gluon plasma  there is a hadronic
gas phase. Using effective Lagrangians we study the interactions of
charmed mesons  which lead to $J/\psi$ production and absorption in this gas. 
We update and extend
previous calculations introducing strange meson interactions and also including
the interactions mediated by the recently measured exotic charmonium resonances
$Z(3900)$ and $Z(4025)$. These resonances open new reaction channels for
the $J/\psi$, which could potentially lead to changes in its multiplicity.
We compute the $J/\psi$ production cross section in processes such as
$D_{(s)}^{(*)} + \bar{D}^{(*)} \to J/\psi + (\pi, \rho, K, K^{\ast})$  and
also the $J/\psi$ absorption cross section in the corresponding inverse
processes.
Using the obtained cross sections as input to solve the rate equation, we
conclude that the interactions in the hadron gas phase  do not
significantly affect the  $J/\Psi$ abundance.
In other words, there is neither a charmonium suppression nor
charmonium regeneration  from  light mesons in the hadronic phase.

\end{abstract}

\maketitle

\section{Introduction}

Precise measurements of the $J/\psi$ multiplicity in heavy ion collisions are
an important source of information about the properties of the quark-gluon
plasma phase \cite{intro}. During this phase $J/\psi$'s are destroyed
and created in a complex and rich dynamical process, which involves many
properties of the QGP which we wish to know better. After cooling and 
hadronization there is a hadron gas phase, which
may  distort or even completely wash out the information  carried by the
$J/\psi$'s about the hot QGP phase. Much work has been devoted to
understand the interactions of the $J/\psi$ in a hadron gas and the most
important process, i.e. the $J/\psi-\pi$ reaction (and the inverse
process), has been exhaustively studied in many papers
\cite{psipi-ft1,psipi-oh,psipi-ft2,psipi-ft3,psipi-qm,psipi-sr,psipi,
psipi-rc,psipi-linn}.
The results of these different calculations  eventually converged
and today we can say that the $J/\psi-\pi$ cross section is known with
a reasonable precision. The $J/\psi$  interactions have
already been  investigated  with field theory models
\cite{psipi-ft1,psipi-oh,psipi-ft2,psipi-ft3}, quark models \cite{psipi-qm},
QCD sum rules \cite{psipi-sr} and other approaches 
\cite{psipi,psipi-rc,psipi-linn}. 
Most of these papers are more than ten years old and
they focus on $J/\psi$ suppression, which had been considered a signature
of the quark-gluon plasma. However, in the last decade experimental
data have shown that at the SPS and at the RHIC  nearly the same  amount
of  $J/\psi$ suppression is observed.  More recently, after the observation
of an ``unsuppression'' at the LHC,
the focus started to be the confirmation of the enhancement of the
$J/\psi$ yield, which became one of the new signatures of the QGP dynamics.

After pions,  kaons are the next lightest and
also very abundant mesons in a hadron gas. 
PHENIX data on particle production in $Au$ - $Au$ collisions \cite{phenix}
show that at low  transverse momentum ($ p_T \simeq  0.5 - 1.5$ GeV) the ratio
$(K^+ + K^-)/(\pi^+ + \pi^-)$ goes to the   value $0.50$.
Recent ALICE data on particle production in $Pb$ - $Pb$ collisions
\cite{alice}
in a similar $p_T$ range show that this ratio is close to  $0.45$.
Particles with these values of $p_T$ most certainly come from the hadron
gas. Taking into account the neutral states, kaons may be up to  30 \% of all
mesons  in the hadron gas. Curiously,  there are quite few
works addressing  the $J/\psi - K$ interaction \cite{regina} and even less
works addressing the $J/\psi - K^*$ interaction \cite{reginas}.
This lack of knowledge and the  potential changes in the final $J/\psi$
abundance that kaons and other strange mesons might cause justifies  the
efforts to  improve  the existing calculations of the  $J/\psi$ - strange
meson dissociation cross sections and also the inverse reactions.          
Indeed, in  his opening talk at 2017 Quark Matter Conference \cite{schu},
J. Schukraft formulated a list of goals to be achieved by the heavy ion
physics community in the near future.  One of them is to understand
processes
such as $D + \bar{D} \to J/\psi + X$,  $D_s + \bar{D} \to J/\psi + X$...etc,
which happen during the late hadronic phase of heavy ion collisions and
increase the number of
$J/\psi$'s. These processes are said to yield ``$J/\psi$ regeneration''
\cite{regen,regen2} and they are a background for  
$J/\psi$ production by recombination of charm-anticharm
pairs during the plasma phase.       

Further motivation to revisit the study of $J/\psi$ interactions with light 
mesons comes from the  striking experimental information which appeared 
after the first round of studies of $J/\psi$ interactions (roughly from 1995  
to 2005): the existence of new charmonium states, the so-called X,Y and Z 
states, which started to be observed in 2003 \cite{exo-rev}. 
Some of these states
generate new  channels for the $J/\psi$-light meson reactions and could 
potentially change the cross sections.  We investige the subject computing 
the cross sections of processes involving $Z_c(3900)$ and $Z_c(4025)$. 

In this work we study  $J/\psi$ production in reactions 
involving pions and strange mesons, such as
$D^{(\ast)} + \bar{D}^{(\ast)} \to J/\psi + \pi $, 
$D^{(\ast)} + \bar{D}^{(\ast)} \to J/\psi + \rho $,
$D_s ^{(\ast)} + \bar{D} ^{(\ast)} \to J/\psi + K $  and
$D_s ^{(\ast)} + \bar{D} ^{(\ast)} \to J/\psi + K^* $. 
As it was pointed out in 
Ref. \cite{psipi-linn} (see Fig. 20 of that work),                  
the reactions initiated by $D$'s and $D_s$'s  have the same order of  
magnitude.  
Making use of the  effective  Lagrangians discussed in 
Refs. \cite{psipi-oh,psipi-ft3,regina,reginas} we will obtain the cross 
sections for the above mentioned processes 
and with them we determine  the thermally averaged cross sections 
for dissociation and production reactions. These latter  are then used as 
input in rate equations, which can be solved giving the $J/\Psi$ abundance in 
heavy ion collisions. In some works (see, for example, Refs. 
\cite{psipi-ft2,Zhou,Ji,Liu}) on the $J/\psi$ dissociation in a hadron gas, 
medium effects are explicitly included. We are not going to take these     
effects into account, since in our formalism the $J/\psi$ interactions are 
already treated  
individually and the use of medium modifications (such as, e.g., in-medium 
masses) might lead to a double counting of the interactions. Also, we are 
not going 
to include in the calculation the $J/\psi$'s which result from the radiative 
decays of the $\psi(2S)$'s.

The paper is organized as follows. In Section~\ref{CrSec} we describe the 
formalism, and determine the production and absorption cross              
sections for  $\pi J/ \Psi, \, \rho J/ \Psi $  and $K ^{(\ast)} J/ \Psi $    
reactions. Then, in Section~\ref{AvCrSec} we present and discuss the results 
obtained for 
thermally averaged cross sections. After that, Section~\ref{TimeEv} is 
devoted to the analysis of $J/\Psi$ abundance in heavy ion               
collisions. Finally, in Section~\ref{Conclusions} we draw the concluding 
remarks.
\section{ Interactions between $J/\Psi $ and light mesons}
\label{CrSec}
 
Our starting point is the calculation of the cross 
sections for the $\varphi -  J/ \Psi$ interactions, 
where $\varphi$ denotes a pseudoscalar or vector  meson. 
To this end,  we follow 
Refs.~\cite{psipi-oh,psipi-ft3,regina,reginas} and use the effective couplings between 
pseudoscalar and vector mesons within the framework of an $SU(4)$ effective theory. 
This is an effective formalism in which the vector mesons are identified as the gauge 
bosons, and the relevant Lagrangians are given by~\cite{psipi-oh,psipi-ft2}
\begin{eqnarray}
\mathcal{L}_{PPV} & = & -ig_{PPV}\langle V^\mu[P,\partial_\mu P]\rangle ,  \nonumber \\
\mathcal{L}_{VVV} & = & i g_{VVV} \langle \partial_\mu V_\nu \left[ V^{\mu}, V^{\nu} 
\right] \rangle , 
\nonumber \\
\mathcal{L}_{PPVV} & = & g_{PPVV}\langle P V^\mu[V_\mu , P]\rangle ,  
\nonumber \\
\mathcal{L}_{VVVV} & = & g_{VVVV}\langle V^\mu V^\nu [V_\mu , V_\nu]\rangle , 
\label{Lagr1}
\end{eqnarray}
where the indices $PPV$ and $VVV$, $PPVV$ and $VVVV$ denote the type of vertex 
incorporating pseudoscalar and vector meson fields in the 
couplings~\cite{psipi-oh,psipi-ft2,regina,reginas}, and 
$g_{PPV}$, $g_{VVV}$, $g_{PPVV}$ 
and $g_{VVVV}$ are the respective coupling constants; the symbol $\langle \ldots \rangle$ 
stands for the trace over $SU(4)$-matrices; $V_\mu$ 
represents a $SU(4)$ matrix, which is parametrized by 16 vector-meson fields including the 
15-plet and singlet of $SU(4)$, 
\begin{eqnarray}
V_\mu = \begin{pmatrix}
\frac{\rho^0}{\sqrt{2}} + \frac{\omega}{\sqrt{6}}  + 
\frac{J /\Psi}{\sqrt{12}} & \rho^+ & K^{*+} & \bar D^{*0} \\
\rho^{-} & - \frac{\rho^0}{\sqrt{2}} + \frac{\omega}{\sqrt{6}}
+\frac{J / \Psi}{\sqrt{12}} & K^{*0} & D^{*-} \\
K^{*-} & \bar K^{*0} & - \frac{2 \omega}{\sqrt{6}} + 
\frac{J /\Psi}{\sqrt{12}} & D^{*-}_s \\
D^{*0} & D^{*+} & D^{*+}_s & - \frac{3 J / \Psi}{\sqrt{12}}
\end{pmatrix}_\mu ;
\label{eq:2}
\end{eqnarray}
$P$ is a matrix containing the 15-plet of the pseudoscalar meson fields, written in the 
physical basis in which $\eta$, $\eta ^{\prime}$ mixing 
is taken into account,
\begin{eqnarray}
P = \begin{pmatrix}
\frac{\pi^0}{\sqrt{2}} + \frac{\eta}{\sqrt{6}}
+\frac{\eta_c}{\sqrt{12}} &
\pi^{+} & K^{+} & \bar D^{0} \\
\pi^{-} & - \frac{\pi^0} {\sqrt{2}} +
\frac{\eta}{\sqrt{6}}+\frac{\eta_c}{\sqrt{12}} & K^{0} & D^{-} \\
K^{*-} & \bar K^{*0} &
- \frac{2 \eta}{\sqrt{6}}+\frac{\eta_c}{\sqrt{12}} & D^{-}_s \\
D^{0} & D^{+} & D^{+}_s & - \frac{3 \eta_c}{\sqrt{12}}
\end{pmatrix} . \nonumber
\end{eqnarray}
In addition to the terms given above, we also consider anomalous parity terms. The 
anomalous parity interactions with vector fields can be 
described in terms of the gauged Wess-Zumino action~\cite{psipi-oh}, which can be 
summarized as
\begin{eqnarray} 
\mathcal{L}_{PVV} & = & - g_{PVV} \varepsilon^{\mu\nu\alpha\beta} \langle 
\partial_\mu V_\nu \partial_\alpha V_\beta P \rangle ,  \nonumber \\
\mathcal{L}_{PPPV} & = & - i g_{PPPV} \varepsilon^{\mu\nu\alpha\beta} 
\langle V_\mu (\partial_{\nu} P) (\partial_{\alpha} P) (\partial_{\beta} P) \rangle , 
\nonumber \\
\mathcal{L}_{PVVV} & = & i g_{PVVV} \varepsilon^{\mu\nu\alpha\beta} \left[  
\langle V_\mu  V_\nu  V_\alpha \partial_{\beta} P \rangle  \right. 
\nonumber \\
& & \left. +  \frac{1}{3} \langle V_\mu (\partial_\nu V_\alpha)  V_\beta P \rangle 
\right].
\label{Lagr2}
\end{eqnarray}
The $g_{PVV}$, $g_{PPPV}$, $g_{PVVV}$ are the coupling constants of the $PVV$, $PPPV$ 
and $PVVV$ vertices, 
respectively~\cite{psipi-oh,psipi-ft2,psipi-ft3,regina,reginas}. 
The couplings given by the effective Lagrangians in Eqs.~(\ref{Lagr1}) 
and~(\ref{Lagr2}) allow us to study the following $\varphi J/ \Psi $ absorption 
processes
\begin{eqnarray}
(1) \;\; \varphi  J/ \Psi & \rightarrow & D_{(s)}  \bar{D}, \nonumber \\
(2) \;\; \varphi  J/ \Psi & \rightarrow &   D_{(s)} ^{\ast} \bar{D} ^{\ast} , 
\nonumber \\
(3) \;\; \varphi J/ \Psi & \rightarrow &   D_{(s)} ^{\ast} \bar{D} , \nonumber \\
(4) \;\; \varphi J/ \Psi &  \rightarrow &  D_{(s)}  \bar{D} ^{\ast} , 
\label{proc1}
\end{eqnarray}
where the final states with strange charmed mesons stand for the initial states 
with $K $ and $K^{\ast} $ mesons, while final states with unflavored charmed mesons 
appear for the initial states with pions and $\rho$ mesons. 
In the present approach, the diagrams considered to compute the amplitudes of the 
processes above are of two types: one-meson exchange and contact graphs.  
They are shown in Fig. 1 of Refs.~\cite{psipi-oh}, \cite{regina} and \cite{reginas} 
for the reactions involving $\pi$, $K$ and $K^{\ast}$, respectively, and in Fig. 2 of
Ref.~\cite{psipi-oh} for those with $\rho$. 

We define the invariant amplitudes for the processes (1)-(4) in Eq. (\ref{proc1}) 
involving $\varphi = \pi, K$ mesons as 
 \begin{eqnarray}
   \mathcal{M}_1 ^{(\varphi)} & = &  \sum _{i} \mathcal{M}_{1i}^{(\varphi)\mu} 
\epsilon_{ \mu} (p_2), \nonumber \\
   \mathcal{M}_2 ^{(\varphi)} & = &  \sum _{i} 
\mathcal{M}_{2i}^{(\varphi)\mu \nu \lambda} \epsilon_{\mu} (p_2) 
\epsilon_{\nu} ^{\ast} (p_3) \epsilon_{\lambda} ^{\ast} (p_4), \nonumber \\
   \mathcal{M}_3 ^{(\varphi)} & = &  \sum _{i} \mathcal{M}_{3i}^{(\varphi)\mu \nu} 
\epsilon_{\mu} (p_2) \epsilon_{\nu} ^{\ast} (p_3) , \nonumber \\
      \mathcal{M}_4 ^{(\varphi)} & = &  \sum _{i} \mathcal{M}_{4i}^{(\varphi)\mu \nu} 
\epsilon_{\mu} (p_2) \epsilon_{\nu} ^{\ast} (p_4) .   
      \label{ampl}
 \end{eqnarray}
 In the above equations, the sum over $i$ represents the sum over all diagrams 
contributing to the respective amplitude; $p_j$  
denotes the momentum of particle $j$, with particles 1 and 2 standing for initial 
state mesons, and particles 3 and 4 for final 
state mesons; $\epsilon_{ \mu} (p_j)$ is the polarization vector related to the 
respective vector particle $j$.   The explicit  expressions of amplitudes 
$\mathcal{M} ^{(\pi)}  $ and $\mathcal{M} ^{(K)}  $ we use in the present work are 
reported in Refs.~\cite{psipi-oh} and \cite{regina}, respectively.

In the case of processes involving $\varphi = \rho, K^{\ast}$ mesons, we must add on 
the  right hand side of each expression in Eq.~(\ref{ampl}) the contraction of the 
amplitude with the polarization vector of vector  meson, i.e. for the reaction 
(1) we have 
$\mathcal{M}_{1}^{(\varphi)\mu \nu} \epsilon_{ \mu} (p_1) \epsilon_{ \nu} (p_2) $
and so on. The explicit expressions of the amplitudes $\mathcal{M} ^{(\rho)}  $  and 
$\mathcal{M} ^{(K^{\ast})} $ used here are those published in Refs.~\cite{psipi-oh} 
and~\cite{regina,reginas}, with some minor changes \cite{correc}.

We are interested in the determination of the isospin-spin-averaged cross section 
for the processes in Eq. (\ref{proc1}), which in the center of mass (CM) frame  is 
defined as
\begin{eqnarray}
  \sigma_r ^{\left(\varphi \right)}(s) 
= \frac{1}{64 \pi^2 s }  \frac{|\vec{p}_{f}|}{|\vec{p}_i|}  \int d \Omega 
\overline{\sum_{S, I}} 
|\mathcal{M}_r  ^{\left(\varphi \right)} (s,\theta)|^2 ,
\label{eq:CrossSection}
\end{eqnarray}
where $r = 1,2,3,4$ labels  $\varphi -  J/ \Psi $ absorption processes according to 
Eq.~(\ref{ampl}); $\sqrt{s}$ is the CM energy;  $|\vec{p}_{i}|$ and $|\vec{p}_{f}|$ 
denote the three-momenta of initial and final particles in the CM frame, respectively; 
the symbol $\overline{\sum_{S,I}}$ represents the sum over the spins and isospins of 
the particles in the initial and final state, weighted by the 
isospin and spin degeneracy 
factors of the two particles forming the initial state for the reaction $r$, i.e. 
\begin{eqnarray}
\overline{\sum_{S,I}}|\mathcal{M}_r|^2 & = & \frac{1}{g_1 g_2 } 
\sum_{S,I}|\mathcal{M}_r|^2, \label{eq:DegeneracyFactors}
\end{eqnarray}
with $g_1 = (2I_{1i,r}+1)(2S_{1i,r}+1), g_2 = (2I_{2i,r}+1)(2S_{2i,r}+1) $ being the 
degeneracy factors of the initial particles 1 and 2. 

We have employed in the computations of the present work the isospin-averaged 
masses:  $m_{\pi} = 138.1 $ MeV, $m_{\rho} = 775.2 $ MeV, $m_K = 495.6 $ MeV, 
$m_{K^{\ast}} = 893.7 $ MeV, $m_D = 1867.2 $ MeV, 
$m_{D^{\ast}} = 2008.6$ MeV, $m_{D_s} = 1968.3 $ MeV, $m_{D_s ^{\ast}} = 2112.1$ 
MeV,  $m_{J/\Psi} = 3096.9 $ MeV. 
Besides, the values of coupling constants appearing in the expressions of the 
amplitudes have been taken from Ref.~\cite{psipi-ft3} 
for $\mathcal{M} ^{(\pi)}  $; from Refs.~
\cite{regina,reginas} for $\mathcal{M} ^{(K)}  $ and $\mathcal{M} ^{(K^{\ast})}  $; 
and from Ref.~\cite{psipi-oh} for the couplings involving $\rho$ meson in 
$\mathcal{M} ^{(\rho)}  $. We have also included form factors in the vertices when 
evaluating the cross sections.  They were taken from \cite{psipi-oh} and are: 
\begin{equation}
	F_3 = \frac{\Lambda^2}{\Lambda ^2 + \mathbf{q} ^2}; \;\;
	F_4 = \frac{\Lambda^2}{\Lambda ^2 + \bar{\mathbf{q}} ^2}
\frac{\Lambda^2}{\Lambda ^2 + \bar{\mathbf{q}} ^2},
\end{equation}
where $F_3$ and $F_4$ are the form factor for the three-point and four-point 
vertices, respectively; $\mathbf{q} =  (\mathbf{p_1}  - \mathbf{p_3}) ^2$ or 
$(\mathbf{p_2}  - \mathbf{p_3}) ^2$ for a vertex involving a $t$- or $u$-channel 
meson exchange; and 
$\bar{\mathbf{q}} = [(\mathbf{p_1}  - \mathbf{p_3}) ^2 + (\mathbf{p_2}  
- \mathbf{p_3}) ^2 ]/2 $.
The cutoff parameter $\Lambda$  was chosen to be $\Lambda = 2.0 $ GeV for all 
vertices~\cite{psipi-oh}.

\subsection{$J/\psi$ absorption}

\begin{figure}[th]
\centering
\includegraphics[width=7.0cm]{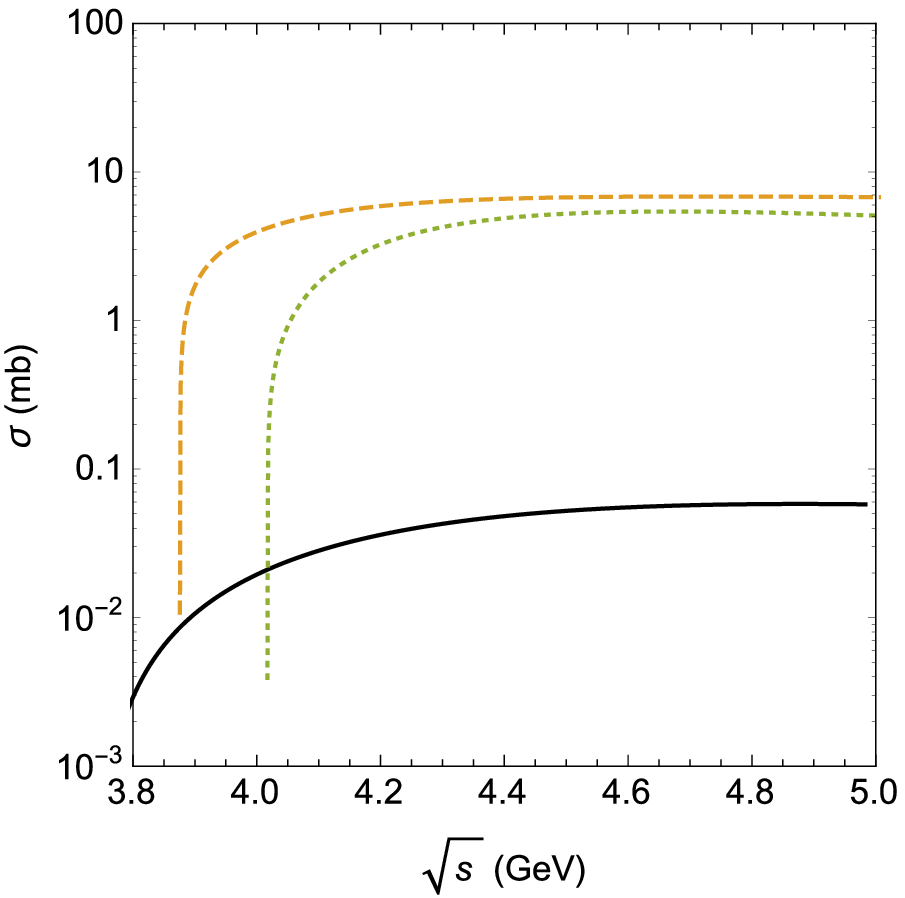}
\hspace{0.5cm}
\includegraphics[width=7.0cm]{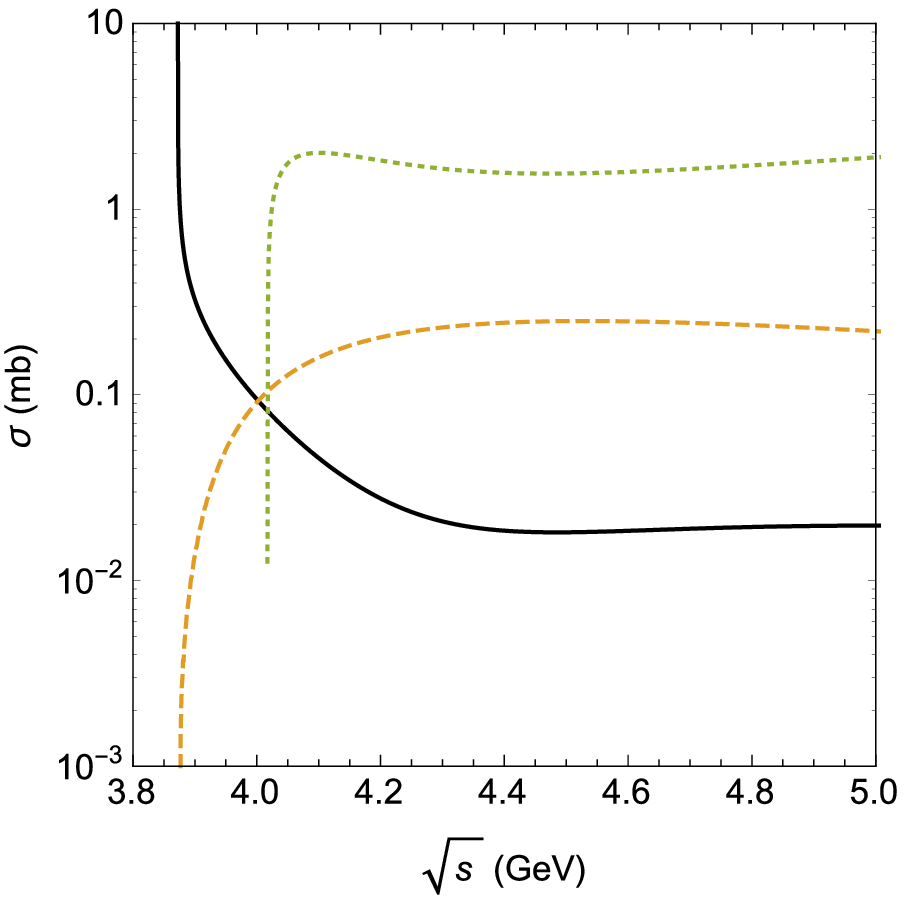} 
\\
\includegraphics[width=7.0cm]{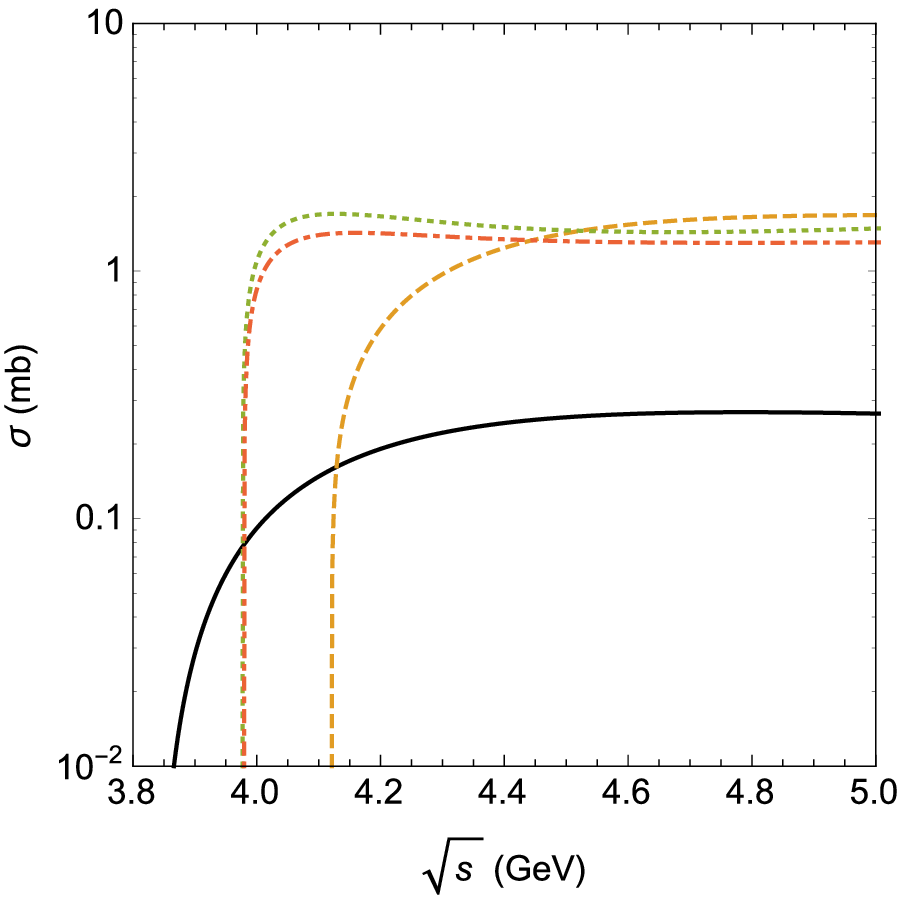}
\hspace{0.5cm}
\includegraphics[width=7.0cm]{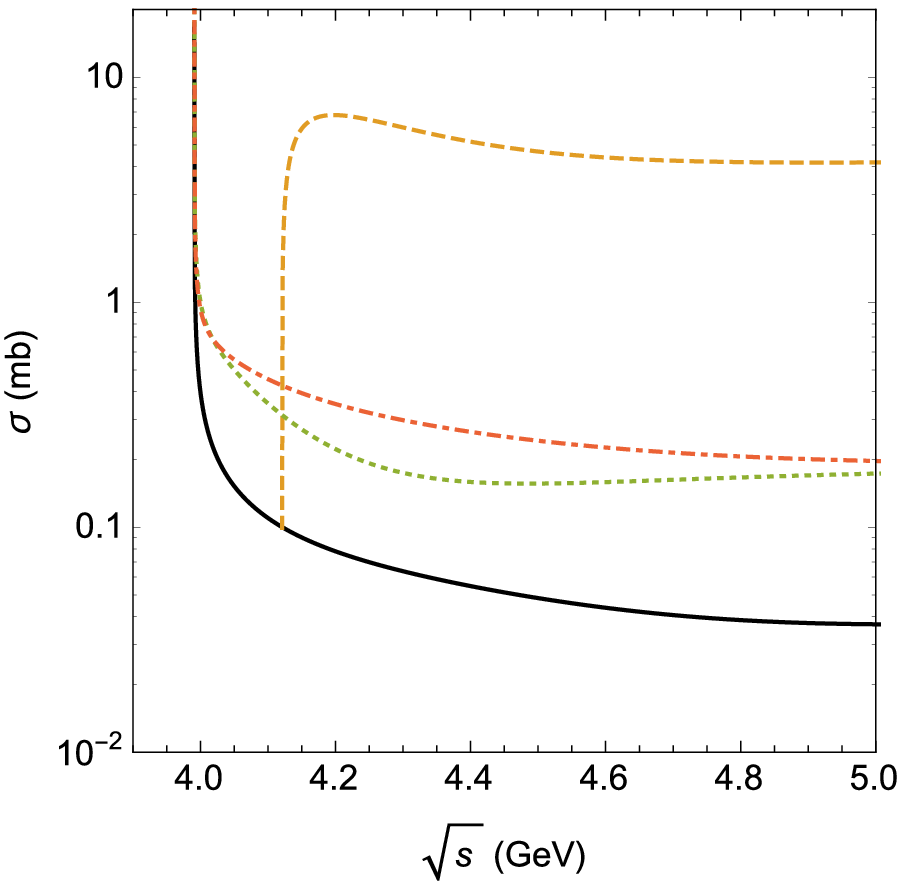}
\caption{$J/\Psi$ absorption cross sections in different processes as a function 
of the CM energy $\sqrt{s}$.  
Top-left panel:   $\pi  J/ \Psi$  in the initial state.   
Top-right panel:  $\rho J/ \Psi$  in the initial state. 
Solid, dashed and dotted lines represent the 
$\pi (\rho) J/ \Psi  \rightarrow  D        \bar{D}       $,  
$\pi (\rho) J/ \Psi  \rightarrow  D^{\ast} \bar{D}       $ and 
$\pi (\rho) J/ \Psi  \rightarrow  D^{\ast} \bar{D}^{\ast}$
reactions, respectively.  
Bottom-left panel:  $K        J/ \Psi$ in the initial state. 
Bottom-right panel: $K^{\ast} J/ \Psi$ in the initial state.
Solid, dashed, dotted and dot-dashed lines represent   the 
$K^{(\ast)} J/ \Psi  \rightarrow  D_s         \bar{D}        $,  
$K^{(\ast)} J/ \Psi  \rightarrow  D_s^{\ast}  \bar{D}^{\ast} $, 
$K^{(\ast)} J/ \Psi  \rightarrow  D_s^{\ast}  \bar{D}        $,   and 
$K^{(\ast)} J/ \Psi  \rightarrow  D_s         \bar{D}^{\ast} $ 
reactions, respectively.}
\label{CrSecJPsi}
\end{figure}

On the top-left panel of Fig.~\ref{CrSecJPsi} the $\pi J/ \Psi $ absorption 
cross sections for the $ \pi J/ \Psi  \rightarrow  D  \bar{D}, D ^{\ast} \bar{D} $ 
and  $D ^{\ast} \bar{D} ^{\ast}$ reactions  are plotted as a function of the CM 
energy $\sqrt{s}$. Both the magnitude and the relative importante of each of these
reactions are in agreement with  previous calculations based on  QCD 
sum rules \cite{psipi-sr}.
The cross sections of the processes  $ \rho J/ \Psi  \rightarrow  D  \bar{D}, 
D ^{\ast} \bar{D} ^{\ast}$ and  $D ^{\ast} \bar{D}^{\ast}$ reactions  are plotted as 
a function of $\sqrt{s}$ on the top-right panel of the figure. We 
see that these cross sections have the same order of magnitude as those 
initiated by pions. This is also in agreemnet with other previous calculations 
(see,  for example, Ref.~\cite{psipi-linn}).
On the bottom-left panel of Fig.~\ref{CrSecJPsi} the  cross sections of the 
processes $K J/ \Psi  \rightarrow  D_s  \bar{D}, D_s ^{\ast} \bar{D} ^{\ast},  
D_s ^{\ast} \bar{D}$, and $ D_s \bar{D} ^{\ast}  $ reactions  are shown. Finally,  
on the bottom-right panel of the same figure we show the cross sections of the  
processes initiated by $K^{\ast}$: $K^{\ast} J/ \Psi  \rightarrow  D_s  \bar{D},  
D_s ^{\ast} \bar{D} ^{\ast},  D_s ^{\ast} \bar{D}$, and $ D_s \bar{D} ^{\ast} $. 
Both the magnitude and the relative importance of each of these reactions are  
in agreement with the results obtained in Ref.~\cite{psipi-linn} and also with those  
obtained in Ref.~\cite{regina} and Ref.~\cite{reginas}. There are some small differences 
due to different choices in the form factors and cutoff values. The most striking 
difference is in the strength of the $K^*$ initiated processes, which in our case 
is remarkably larger. 

	
Summarizing, despite the different $ \sqrt{s}$-dependence of the 
$(\pi,\rho,K , K^{\ast}) - J/ \Psi $ absorption cross sections discussed above, 
their contributions can be considered approximately of the same order of magnitude, 
justifying the inclusion of all these contributions in the analysis of $J/\Psi$ 
abundance that will be done in next Sections.  

\subsection{$J/\psi$ production} 

\begin{widetext}

\begin{figure}[th]
\centering
\includegraphics[width=7.0cm]{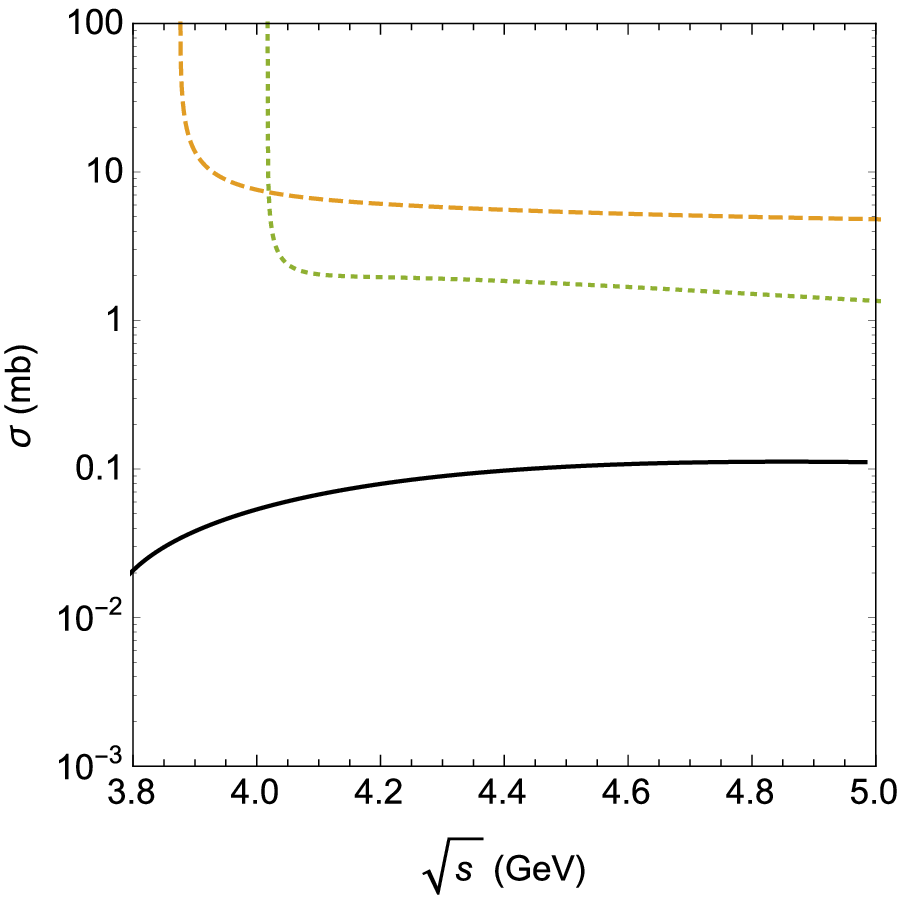}
\hspace{0.5cm}
\includegraphics[width=7.0cm]{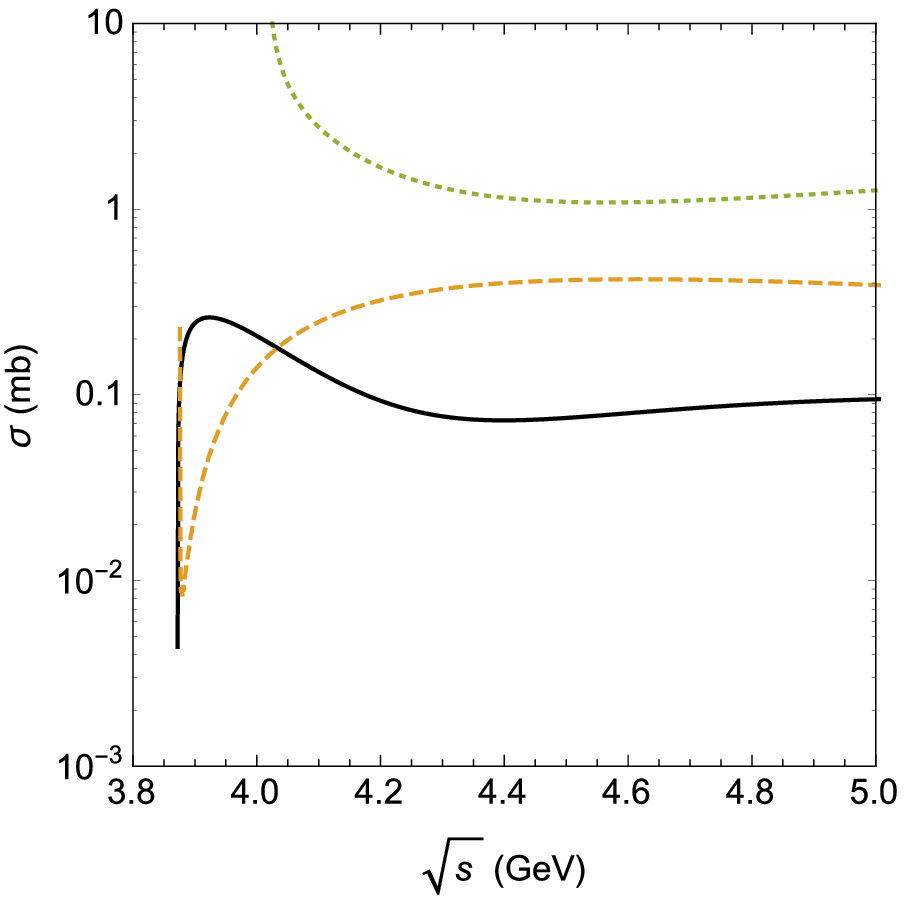}
\\
\includegraphics[width=7.0cm]{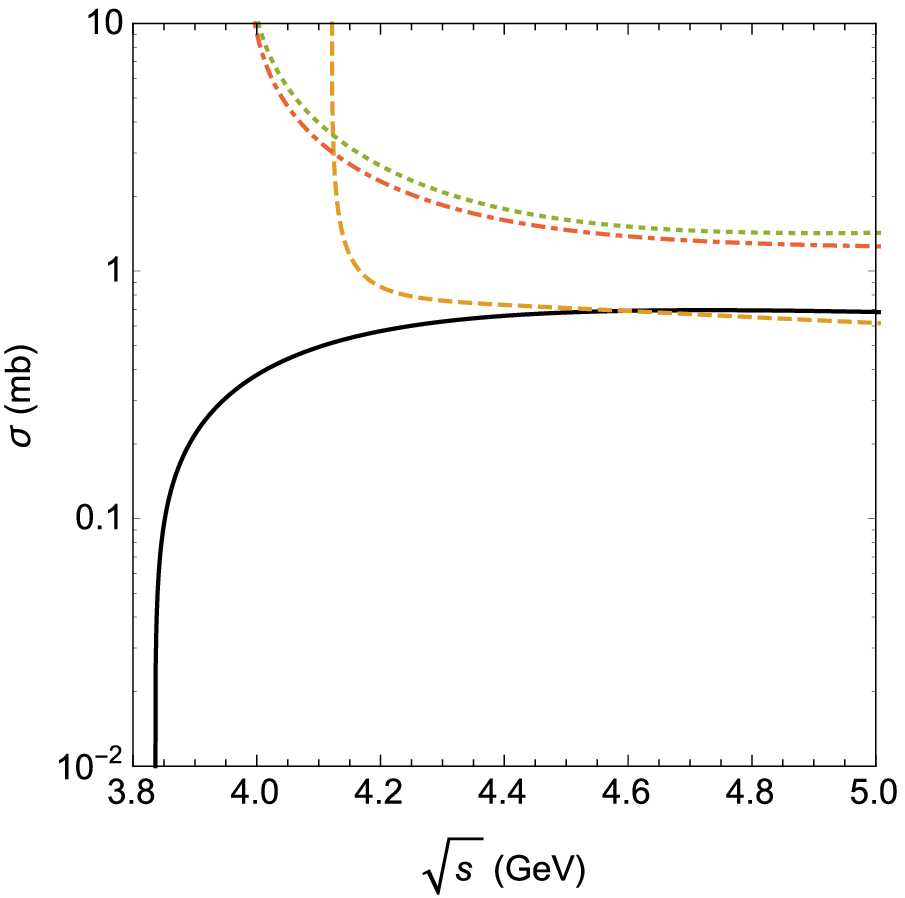}
\hspace{0.5cm}
\includegraphics[width=7.0cm]{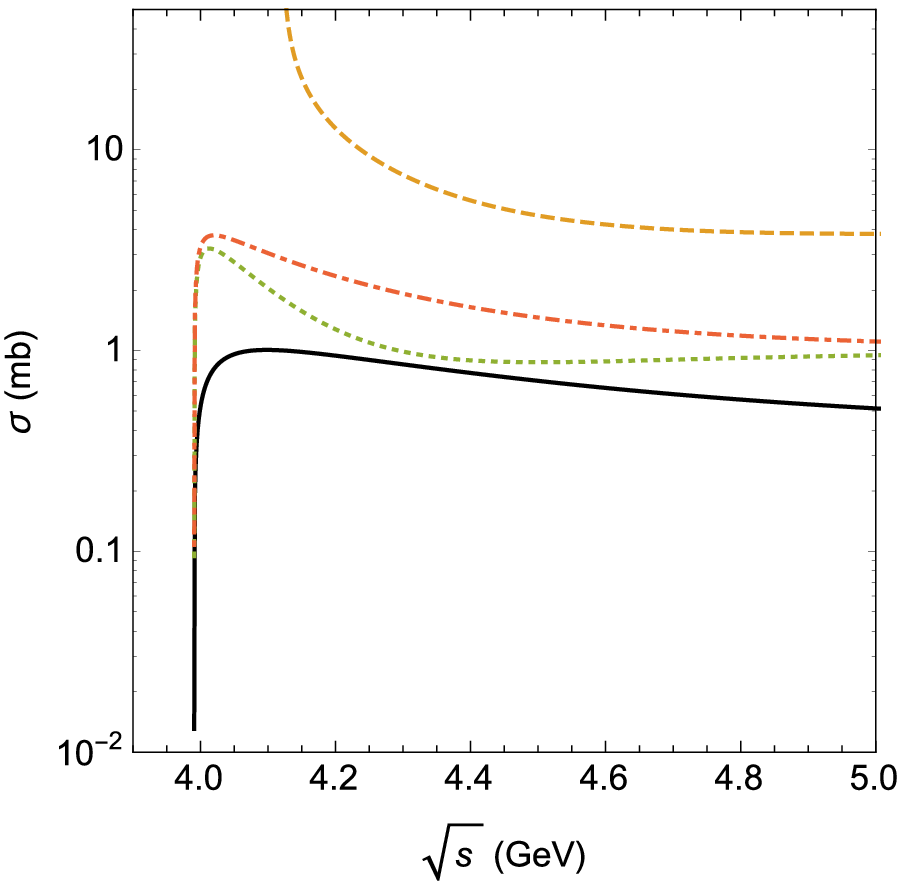}                  
\caption{$J/\Psi$ production cross sections in different processes as a 
function  of the CM energy $\sqrt{s}$.
Top-left panel:  $\pi  J/ \Psi$ in the final state. 
Top-right panel: $\rho J/ \Psi$ in the final state.
Solid, dashed and dotted lines represent the 
$ D         \bar{D}         \rightarrow \pi (\rho) J/ \Psi $, 
$ D^{\ast}  \bar{D}         \rightarrow \pi (\rho) J/ \Psi $ and 
$ D^{\ast}  \bar{D}^{\ast}  \rightarrow \pi (\rho) J/ \Psi $ 
reactions, respectively. 
Bottom-left panel:  $K        J/ \Psi$ in the final state. 
Bottom-right panel: $K^{\ast} J/ \Psi$ in the final state.
Solid, dashed, dotted and dot-dashed lines represent the 
$ D_s        \bar{D}         \rightarrow K^{(\ast)} J/ \Psi $, 
$ D_s^{\ast} \bar{D}^{\ast}  \rightarrow K^{(\ast)} J/ \Psi $,  
$ D_s^{\ast} \bar{D}         \rightarrow K^{(\ast)} J/ \Psi $, and 
$ D_s        \bar{D}^{\ast}  \rightarrow K^{(\ast)} J/ \Psi$ 
reactions, respectively.}
\label{CrSecJPsiInv}
\end{figure}
\end{widetext}

We now calculate the cross sections of the inverse processes, which can be 
obtained from the direct processes through the use of detailed balance 
(see Eq.~(48) from Ref.~\cite{psipi-ft3}). 
In the top-left panel of 
Fig.~\ref{CrSecJPsiInv} the $\pi J/ \Psi $ production cross sections for the $D  
\bar{D}  \rightarrow  \pi J/ \Psi,  D ^{\ast} \bar{D} ^{\ast}  \rightarrow  
\pi J/ \Psi$ and $  D ^{\ast}  \bar{D}  \rightarrow  \pi J/ \Psi $  reactions  
are plotted as a function of the CM energy $\sqrt{s}$. 
In the  top-right panel of Fig.~\ref{CrSecJPsiInv} the $\rho J/ \Psi $ production 
cross sections for the $D  \bar{D}  \rightarrow  \rho J/ \Psi,  
D ^{\ast} \bar{D} ^{\ast}  \rightarrow  \rho J/ \Psi$ and 
$  D ^{\ast}  \bar{D}  \rightarrow  \rho J/ \Psi $  reactions  are plotted as a 
function of the CM energy $\sqrt{s}$. 
In  the bottom-left  panel of Fig.~\ref{CrSecJPsiInv} the $K J/ \Psi $ production  
cross sections are plotted and in  the  bottom-right  panel of same figure the  
$K^{\ast} J/ \Psi $ production cross sections are plotted. 

\begin{figure}[t]
  \begin{tabular}{cc}
   {\psfig{figure=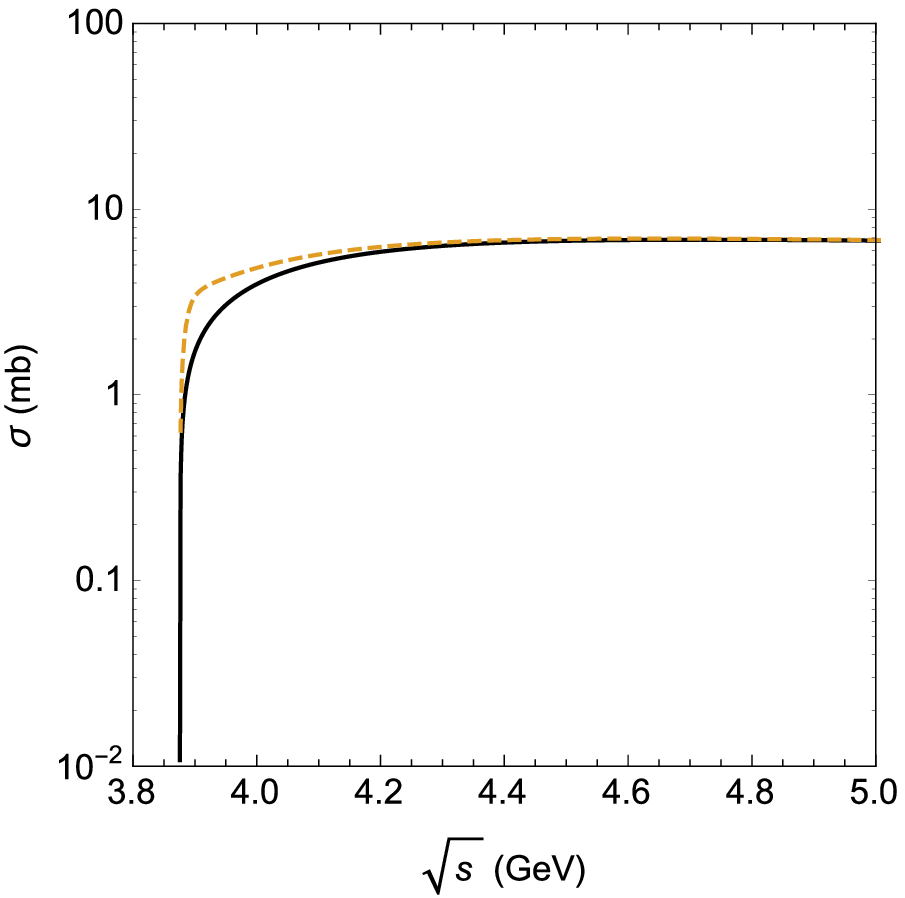,scale=0.82}} &
   {\psfig{figure=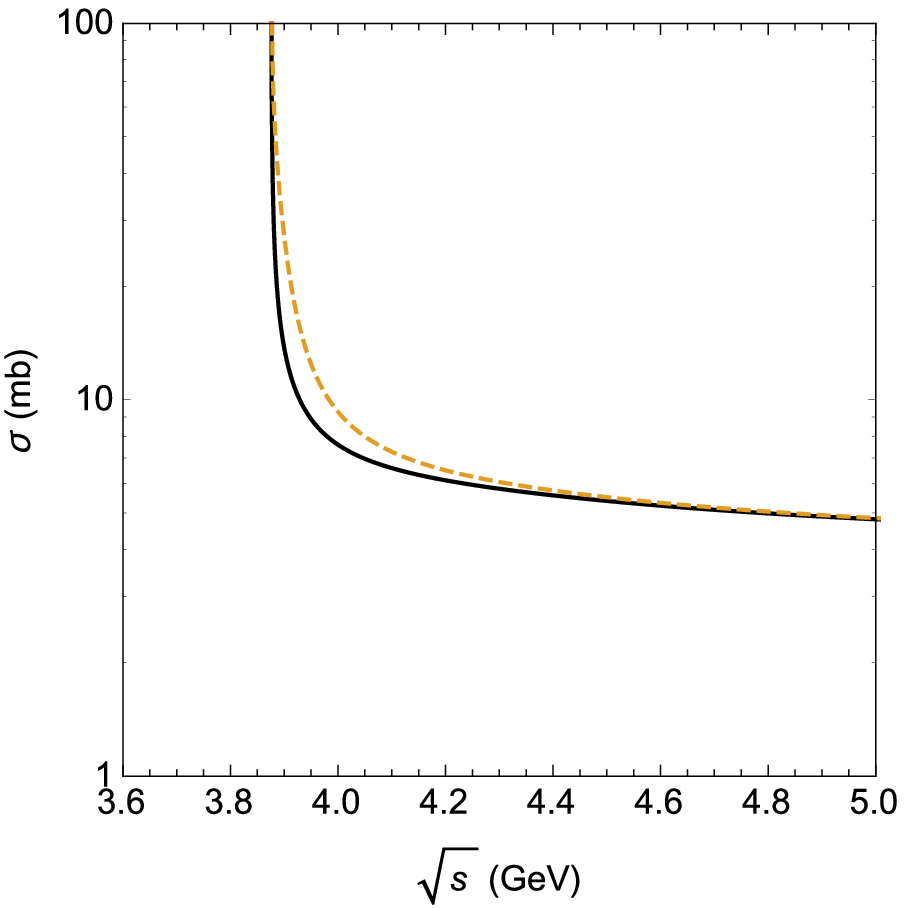,scale=0.82}}  \\
   {\psfig{figure=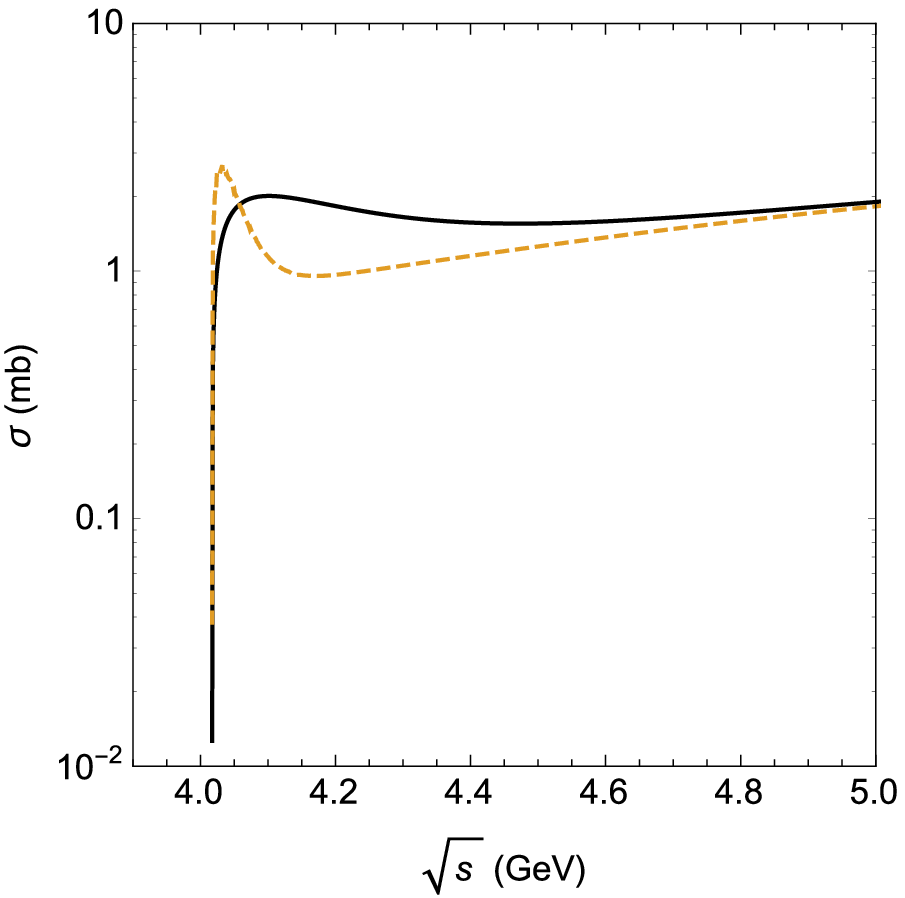,scale=0.82}}
 & {\psfig{figure=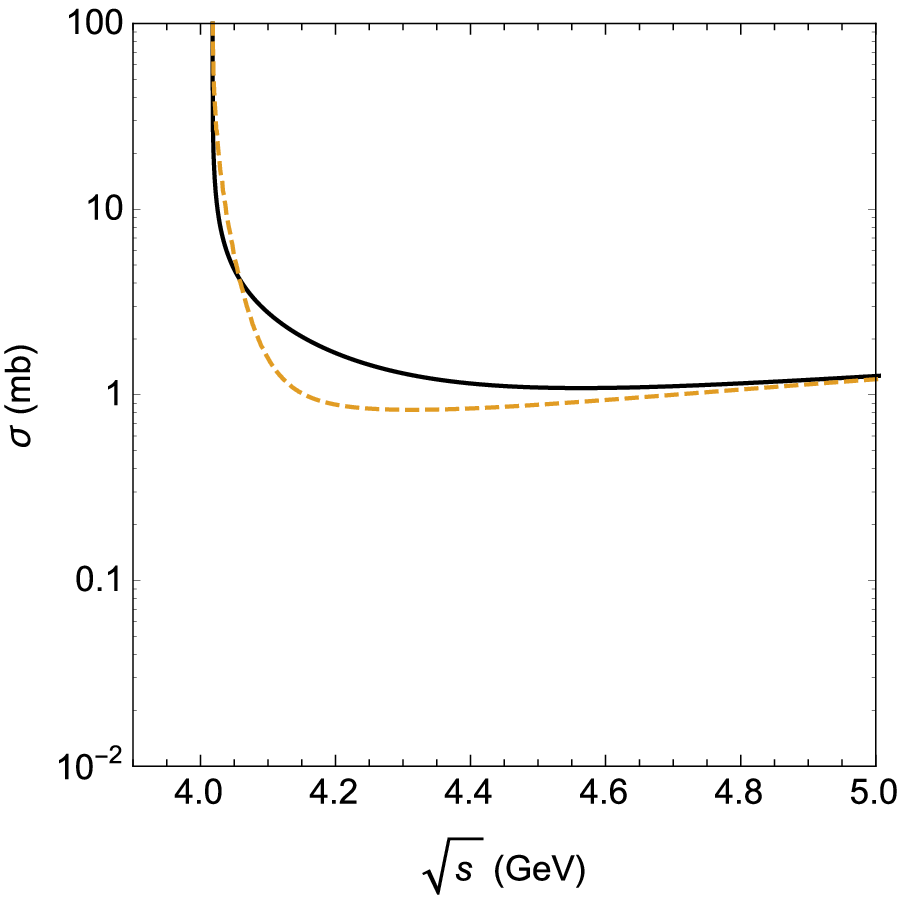,scale=0.82}}
  \end{tabular}
\caption{$J/\Psi$ absorption (top left) and production (top right ) cross sections  
by $\pi$'s. The solid lines represent the cross sections obtained without including
the $Z_c$ (3900) exchange in the s-channel. The dashed lines show the results with  
the exchange of $Z_c$ (3900) in the s-channel included. Bottom panels show the   
$J/\Psi$ absorption (bottom left ) and production (bottom right ) cross sections   
by $\rho$’s. The solid lines in these panels show the cross sections obtained   
without  including the $Z_c$ (4025) exchange in the s-channel. The dashed lines 
show the results obtained by including the $Z_c$(4025) exchange in the s-channel.
}                                                           
\label{zcpi} 
\end{figure}

\begin{widetext}

\end{widetext}

\begin{widetext}          
\begin{figure}[th] 
\centering 
\includegraphics[width=7.0cm]{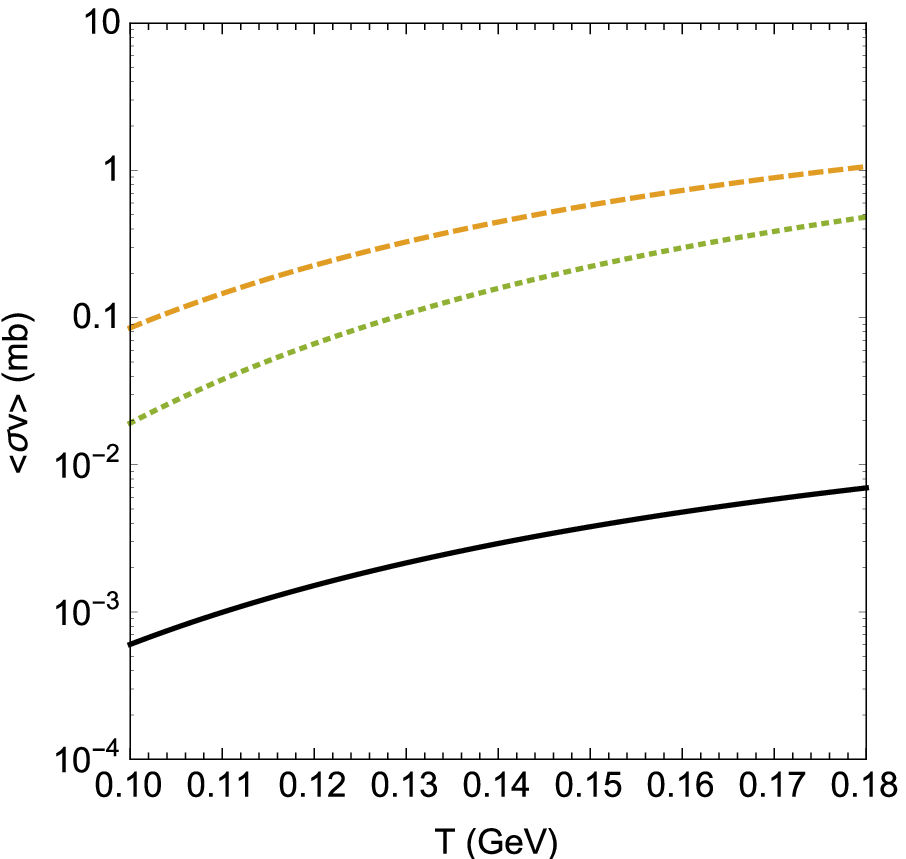}
\hspace{0.6cm} 
\includegraphics[width=7.0cm]{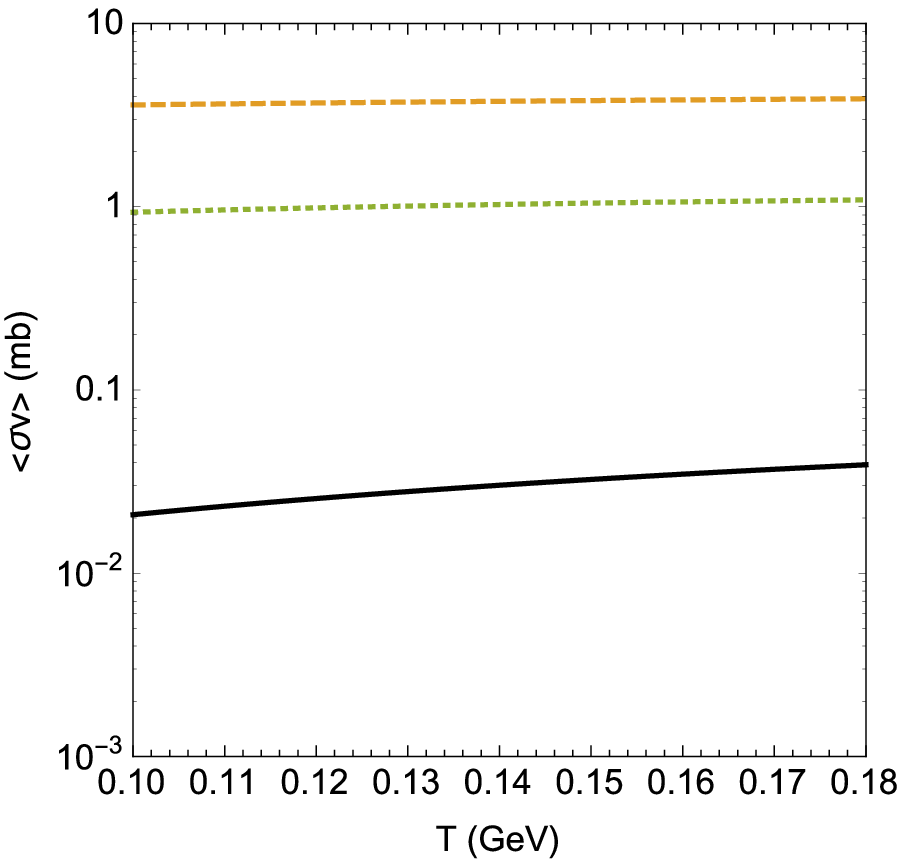}
\\ 
\includegraphics[width=7.0cm]{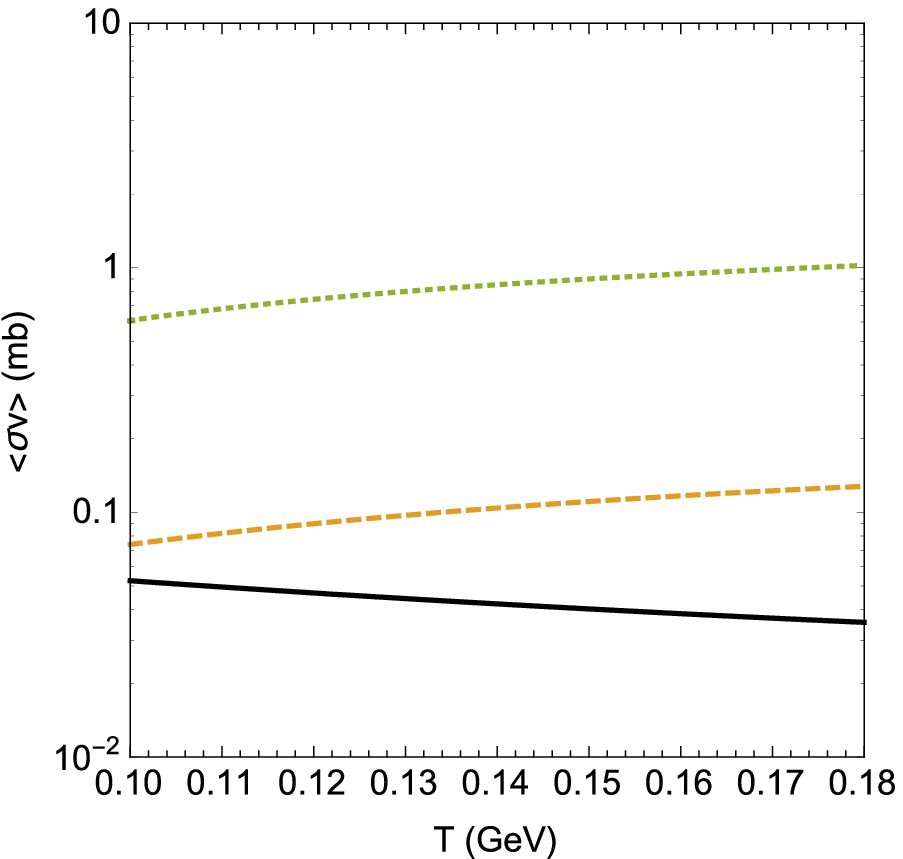}
\hspace{0.6cm} 
\includegraphics[width=7.0cm]{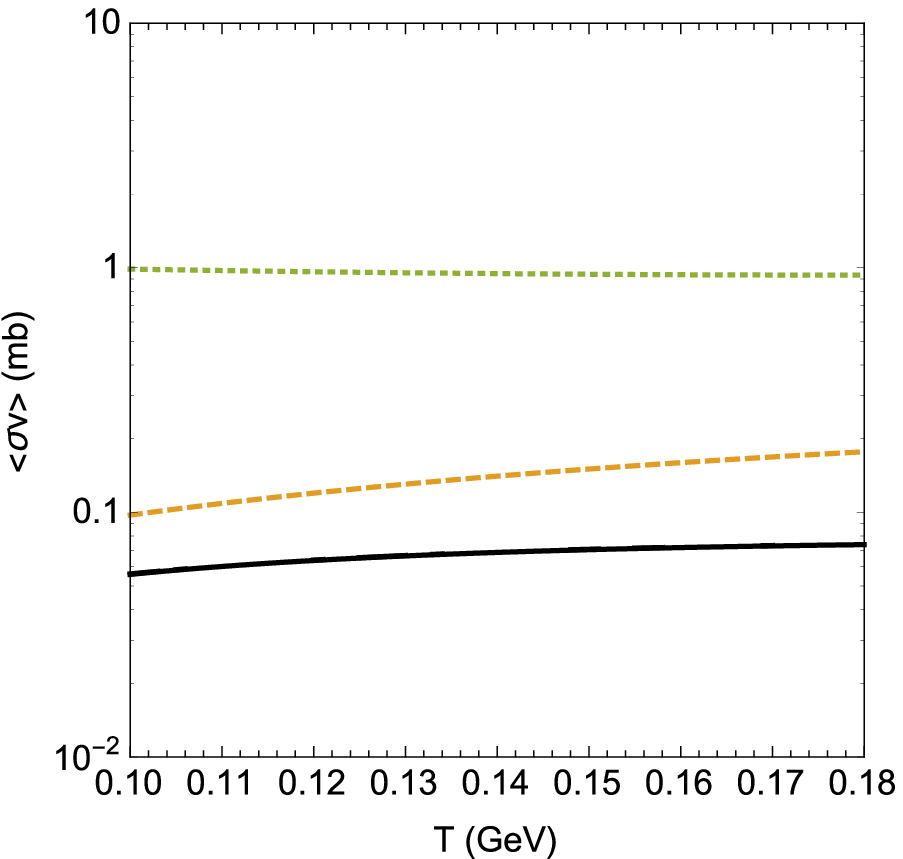}
\caption{$J/\Psi$ absorption and production cross sections 
by $\pi$'s and $\rho$'s as a  function  of the temperature. 
Top-left panel:  absorption reactions with $\pi  J/ \Psi$ in the initial state.  
$\pi  J/ \Psi   \rightarrow D  \bar{D} $ (solid line),  
$\pi  J/ \Psi   \rightarrow D^{\ast}  \bar{D}  $ (dashed lines) and 
$\pi  J/ \Psi   \rightarrow D^{\ast}  \bar{D}^{\ast}  $ (dotted lines).  
Top-right panel:  production reactions with $\pi  J/ \Psi $ in the final state.
The line convention is the same as in the left panel.
Bottom-left panel: absorption reactions with  $\rho J/ \Psi$ in the initial state.   
$\rho  J/ \Psi  \rightarrow  D        \bar{D}        $ (solid line),  
$\rho  J/ \Psi  \rightarrow  D^{\ast} \bar{D}        $ (dashed line) and    
$\rho  J/ \Psi  \rightarrow  D^{\ast} \bar{D}^{\ast} $ (dotted line)
Bottom-right panel: production reactions with $\rho J/ \Psi$ in the final state.  
The line convention is the same as in the left panel.
} 
\label{thermal-light} 
\end{figure} 
\end{widetext} 

From these many curves, two general conclusions may be drawn: i) Reactions 
which start or end with $\pi J/\Psi$ and $K^* J/\Psi$ have larger cross sections. 
ii) Excluding the low energy region (which will be much less relevant for 
phenomenology), the $J/\psi$ production and absorption cross sections are very 
close to each other in almost all channels. 
Since the $J/\Psi$ absorption and production cross sections have 
comparable magnitudes, what will determine the final yield of $J/\Psi$'s will 
be the thermally averaged cross sections, which, reflecting the physical aspects 
of the hadron gas, will select the range of energies (in the horizontal axis of 
Figs. \ref{CrSecJPsi}  and \ref{CrSecJPsiInv}) which are more important. 

\subsection{The impact of the $Z(3900)$ and $Z(4025)$ resonances on  $J/\psi$ 
production}

Over the last decade the existence of exotic charmonium states has been well 
established. These are new states which contain a $c \bar{c}$ pair but are not
conventional  quark-antiquark configurations, being rather multiquark states.
For the present work some states  are particularly 
relevant: those which decay into $J/\psi -\pi$, as the $Z_c(3900)$, and
those which decay into $J/\psi-\rho$, as the $Z_c(4025)$.  
Indeed, these  states open  new s-channels for $J/\psi$ interactions, 
as, for example:  
$J/\psi +  \pi \to  Z_c \to D + \bar{D}^* $. These processes can change the 
results found in the previous section and hence deserve a special attention. 
The impact of the best known of the exotic states, the $X(3872)$, on $J/\psi$ 
interactions with light mesons was first investigated in Ref. \cite{brazzi}, 
where the cross section of the reaction 
$J/\psi + \rho  \to D + \bar{D}^*$  was calculated.  The obtained cross section 
was very small and can then be neglected. More recently the BES Collaboration 
observed \cite{bes} and then confirmed \cite{bes3} the $Z^\pm_c(3900)$  state
in the $\pi \, J/\psi$ invariant mass of the reaction 
$e^+ e^-\to \pi^+\pi^-J/\psi$, with $J^P = 1^+$, mass  $3881.2\pm 4.2\pm 52.7$ 
MeV and  width $51.8\pm 4.6\pm 36$ MeV. Signals for its neutral partner, 
$Z^0_c(3900)$, have 
also been found \cite{bes4}. In the context of the present work, a natural 
question is then: what is the impact of the reaction 
$ J/\psi + \pi \to Z_c \to D + \bar{D}^*$ on the results found in the previous 
section ?  In order to estimate the influence of the $Z_c(3900)$ on such 
reactions, we consider the process of the absorption of $J/\psi$ for a 
particular total electric charge, which can be 0, $+1$ or $-1$. The amplitude 
for this process can be written as
\ben
\mathcal{M}_{Z} &=&
\alpha_{J/\psi \pi} \, \alpha_{D\bar D^*}\frac{1}{s-M^2_Z+i M_Z\Gamma_Z} 
\nonumber \\
&& \times \, \left(-g^{\mu\nu}+\frac{p^\mu k^{\,\prime\nu}}{M^2_Z}\right)
\epsilon^\mu(k) \epsilon^{\nu\,*}(p^\prime) \, , 
\label{MZ}
\een
where $M_Z = 3871.28$ MeV and $\Gamma_Z = 40$ MeV represent the mass and width  
of the $Z_c(3900)$ respectively. Also,   $\alpha_{J/\psi\pi}$ and    
$\alpha_{D\bar D^*}$ are the couplings of the $Z_c$ to the $J/\psi \pi$ and to           
the $D\bar D^{*}$ states, respectively, for a particular electric charge. 
To determine    
these couplings we use the results of Ref.~\cite{aceti}.
In Table~\ref{couplings} we show the values found for these couplings for the 
channels of interest in this work. In the Table, the
quantities $\alpha_1=8128.3-i\,53.0$ MeV and $\alpha_2=3300-i\,923$ MeV
represent the coupling of the $Z_c$ to the states
$\frac{1}{\sqrt{2}}(|D\bar D^*,I=1\rangle+ |D^*\bar D,I=1\rangle)$
and $|J/\psi\pi\rangle$, which have isospin 1 and positive $G$-parity. 
We follow  the isospin phase convention $|\pi^+\rangle=-|1,1\rangle$,
$|D^0\rangle=|D^{*0}\rangle=-|1/2,-1/2\rangle$.
\begin{table}[h!]
\centering
\caption{Couplings of the $Z_c(3900)$ found in Ref.~\cite{aceti} to the   
channels $J/\psi\pi$ and $D\bar D^*$ for different electric charges.}
\label{couplings}
\begin{tabular}{c|c}
\text{Channel}&$\text{Coupling}$\\
\hline\\
$D^0\bar D^{*0}$, $D^+ D^{*-}$& $-\alpha_1/2$, $\alpha_1/2$\\               
$D^+\bar D^{*0}$, $D^0 D^{*-}$& $\alpha_1/\sqrt{2}$, -$\alpha_1/\sqrt{2}$\\
$J/\psi \pi^{0}$, $J/\psi \pi^{-}$, $J/\psi \pi^{+}$& $\alpha_2$, 
$\alpha_2$, -$\alpha_2$
\end{tabular}
\end{table}

The BESIII collaboration has also claimed the existence of an isospin 1  
resonance, called $Z_c(4025)$ (width around 25 MeV) in the $D^*\bar D^*$   
invariant mass distribution of the reaction  
$e^+ e^-\to \pi^\mp (D^*\bar D^*)^\pm$~\cite{bes5}.                       
Assuming  that the $D^*\bar D^*$ pair interacts in s-wave, the authors of  
Ref.~\cite{bes5} have assigned to the  $Z_c(4025)$ the quantum 
numbers $J^P=1^+$.   
However, as stated by the same authors, the experiment can not exclude other 
spin-parity assignments.
In fact, as shown in Ref.~\cite{mknno2}, the invariant mass distribution  
found in Ref.~\cite{bes5} can be explained considering the                   
$I^G(J^{PC})=1^- (2^{++})$ state with mass and width around 4000 and 90 MeV,  
respectively, which is generated  as a consequence of the interaction of 
$D^*\bar D^*-\text{c.c}$ and $J/\psi \rho$ in a coupled channel 
approach~\cite{mo,acetibjo,kmnn} . This interpretation is more plausible, 
since if the state $Z_c(4025)$ would have $J^P=1^+$, as assumed in          
Ref.~\cite{bes5}, it should have a large decay width to $J/\psi \pi$, as in   
case of the $Z_c(3900)$ mentioned above. However, in Refs.~                 
\cite{bes,belle,cleo,bes5} no signal is found in the $J/\psi \pi$ invariant  
mass around 4025 MeV. Note that, theoretically, in both cases, the states  
$Z_c(3900)$ and $Z_c(4025)$ appear below the $D\bar D^*$ threshold for the  
former and below the $D^*\bar D^*$ threshold for the latter. However, as 
explained in Ref.~\cite{mknno2}, it is their corresponding widths what 
makes possible their manifestation in the $D\bar D^*$ and $D^*\bar D^*$ 
invariant mass distribution found in Refs.~\cite{bes,belle,cleo,bes5}.

As in the case of the $Z_c(3900)$, the exchange of $Z_c(4025)$ could also play an 
important role in the determination of  the cross section of the reaction        
$J/\psi \rho \to D^*\bar D^*$ and its time reversed process. In order to 
estimate this contribution we consider the s-channel process
$ J/\psi + \rho \to Z_c(4025) \to D^* + \bar{D}^*$ and 
follow Ref.~\cite{acetibjo}, where $Z_c(4025)$ is associated with the       
$I^G(J^{PC})=1^+(2^{++})$ state found at 3998 MeV with a width of 90 MeV in 
the $T$-matrix obtained from the resolution of the Bethe-Salpeter equation 
considering $D^*\bar D^*-\text{c.c}$, $J/\psi\rho$ as coupled channels. 
The amplitude associated with the process
$ J/\psi + \rho \to Z_c(4025) \to D^* + \bar{D}^*$  is given by
\ben
\mathcal{M}_{Z^\prime} &=& 
\eta_{J/\psi\rho} \, \eta_{D^*\bar D^*}\frac{1}{s-M^2_{Z^\prime}+
i M_{Z^\prime}\Gamma_{Z^\prime}}  \nonumber \\
&&
\times \, P^{\mu\nu\alpha\beta}(q)
\epsilon_\mu(k)\epsilon_\nu(p)\epsilon^*_\alpha(k^\prime)
\epsilon^*_\beta (p^\prime),
\een
where $M_{Z^\prime}=3989.61$ MeV and $\Gamma_{Z^\prime}=90$ MeV are,
respectively, the mass and width found for the $Z_c(4025)$ in
Ref.~\cite{acetibjo}, $q=k+p=k^\prime+p^\prime$ is the total
four-momentum, $\eta_{J/\psi\rho}$ and $\eta_{D^*\bar D^*}$ are the
couplings of $Z_c(4025)$ to the channels $J/\psi \rho$ and $D^*\bar D^*$
for a particular total electric charge (0, $+1$ or $-1$) and
$P^{\mu\nu\alpha\beta}(q)$ is the spin 2 projector, which is given
by~\cite{kmnn,mknno3}
\begin{equation}
P^{\mu\nu\alpha\beta}(q)=\frac{1}{2}(\Delta^{\mu\alpha}\Delta^{\nu\beta}+
\Delta^{\mu\beta}\Delta^{\nu\alpha})-\frac{1}{3}\Delta^{\mu\nu}
\Delta^{\alpha\beta},
\end{equation}
with
\begin{equation}
\Delta^{\mu\nu}(q)=-g^{\mu\nu}+\frac{q^\mu q^\nu}{M^2_{Z^\prime}},
\end{equation}
and $g^{\mu\nu}=\text{diag}(1~-1~-1~-1)$ is the metric tensor.  
\begin{table}[h!]
\centering
\caption{Coupling of the $Z_c(4025)$ found in Ref.~\cite{acetibjo} to the  
channels  $J/\psi\rho$ and $D^*\bar D^*$ for different electric charges.}
\label{eta}
\begin{tabular}{c|c}
\text{Channel}&$\text{Coupling}$\\
\hline\\
$D^{*0}\bar D^{*0}$, $D^{*+} D^{*-}$& $-\eta_1/2$, $\eta_1/2$\\
$D^{*+}\bar D^{*0}$, $D^{*0} D^{*-}$& $\eta_1/\sqrt{2}$, -$\eta_1/\sqrt{2}$\\ 
$J/\psi \rho^{0}$, $J/\psi \rho^{-}$, $J/\psi \rho^{+}$& $\eta_2$, $\eta_2$, 
-$\eta_2$
\end{tabular}
\end{table}
The coupling constants $\eta_{J/\psi\rho}$ and $\eta_{D^*\bar D^*}$, as in    
case of  $Z_c(3900)$, can be calculated from the residue of the corresponding  
scattering matrix of Ref.~\cite{acetibjo} evaluated at the pole position. The 
value for these couplings are listed in Table~\ref{eta}. In the Table 
the quantities $\eta_1=12560.80-i\,507.80$ MeV and $\eta_2=8145.78-i\,2627.96$
MeV represent the coupling of $Z_c(4025)$ to the states
$\frac{1}{\sqrt{2}}(|D \bar D^*,I=1\rangle- |D^*\bar D,I=1\rangle)$
and $|J/\psi\rho\rangle$, which have isospin 1 and $G$-parity
negative. We follow the isospin phase convention $|\rho^+\rangle=-|1,1\rangle$,
$|D^{*0}\rangle=-|1/2,-1/2\rangle$.  With the above amplitudes we can calculate 
the cross sections for $J/\psi$ absorption and production in each one of the 
channels.  

In Fig.~\ref{zcpi} we show the cross sections of the  processes  
$J/\psi + \pi \to D + \bar{D}^*$  and $J/\psi + \rho \to D + \bar{D}^*$  
and the corresponding inverse processes. The solid lines show the 
results obtained in the previous subsections and the dashed lines show the 
effect of including the $Z_c(3900)$ and $Z_c(4025)$ as described above. 
As it can be seen the effect of the new resonances is  small and will be 
neglected  in what follows.

\section{Thermally averaged cross sections}
\label{AvCrSec}
We define the thermally averaged cross section for a given process 
$a b \rightarrow c d$ as~\cite{Koch,Cho:2010db,Cho:2015qca,Abreu:2016qci}
\ben
\langle \sigma_{a b \rightarrow c d } \, v_{a b}\rangle &  = & 
\frac{ d^{3} \mathbf{p}_a \, d^{3} \mathbf{p}_b \,  
f_a(\mathbf{p}_a) \,  f_b(\mathbf{p}_b) \,  \sigma_{a b \rightarrow c d } \,\,v_{a b} }
{ d^{3} \mathbf{p}_a \,   d^{3} \mathbf{p}_b \,  
f_a(\mathbf{p}_a) \, f_b(\mathbf{p}_b) } 
\label{thermavcs}
\een
where $v_{ab}$ represents the relative velocity of initial two interacting 
particles $a$ and $b$ and the function $f_i(\mathbf{p}_i)$ is the Bose-Einstein 
distribution (of particles of species i), which depends on the temperature $T$.

In the two upper panels of Fig.~\ref{thermal-light} we plot the thermally 
averaged cross sections for  $\pi J/ \Psi $ absorption (on the left) and 
production (on the right) via the 
processes discussed in previous section. We can see that  for all processes     
the production reactions are larger than the absorption ones.  
In the  two lower panels of Fig.~\ref{thermal-light} we plot the thermally 
averaged cross sections for the $\rho J/ \Psi $ absorption and production. It can 
be noticed that they are comparable for all processes. 
\begin{widetext}           
 \begin{figure}[th]  
\centering  
\includegraphics[width=7.0cm]{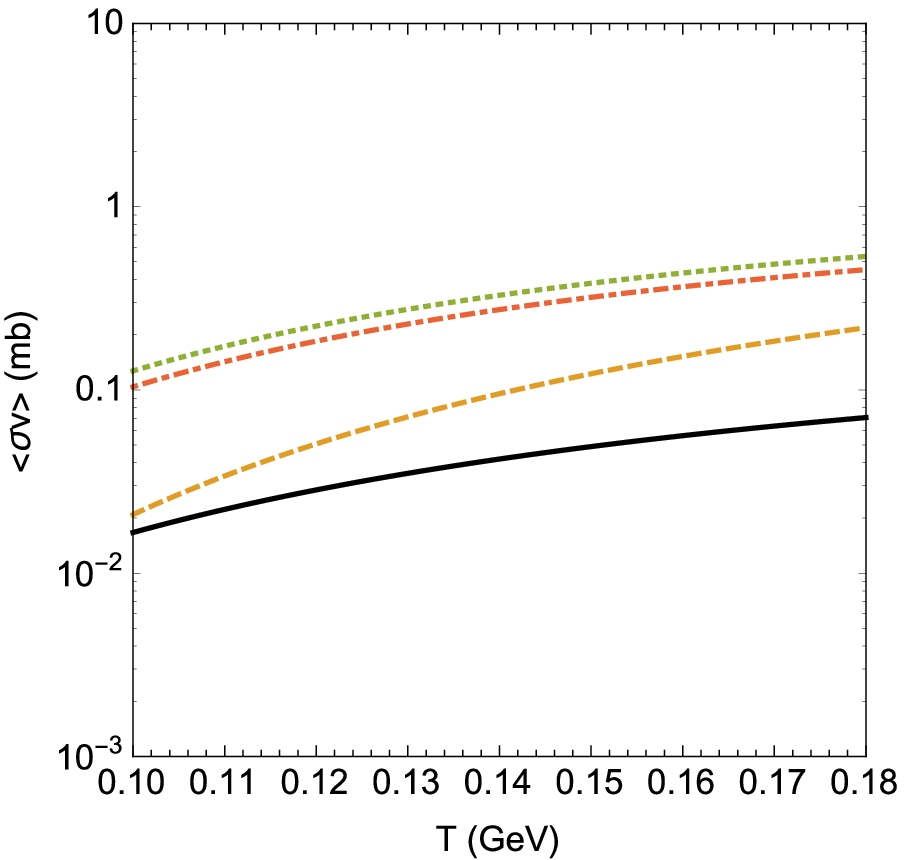}  
\hspace{0.6cm}  
\includegraphics[width=7.0cm]{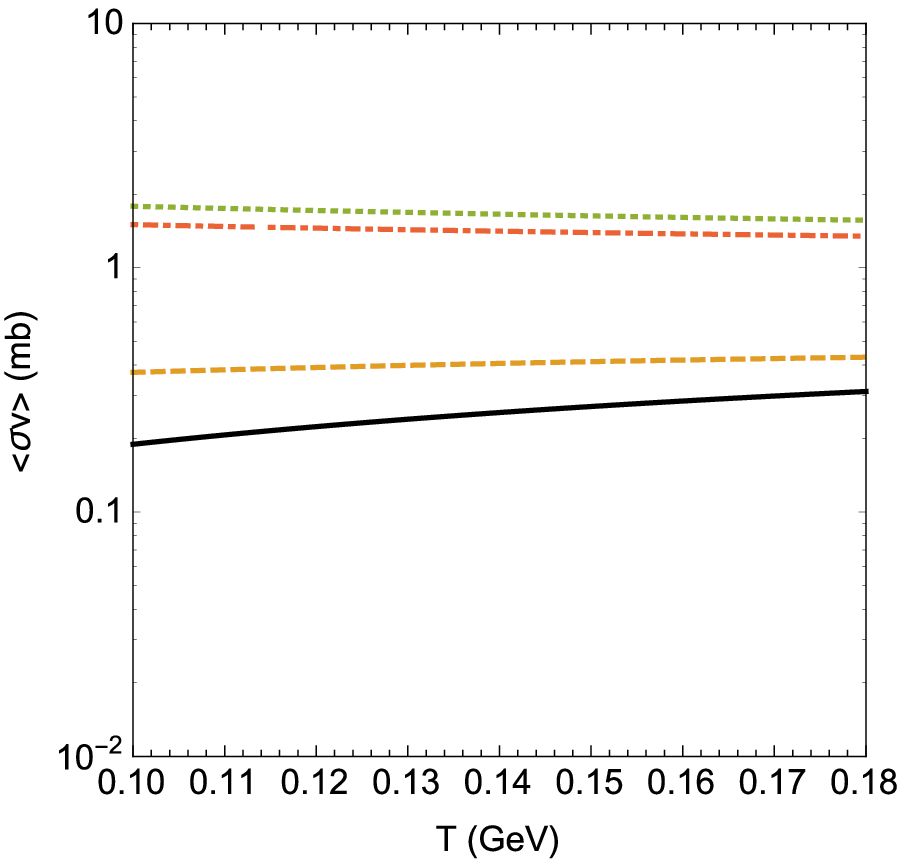}
\\  
\includegraphics[width=7.0cm]{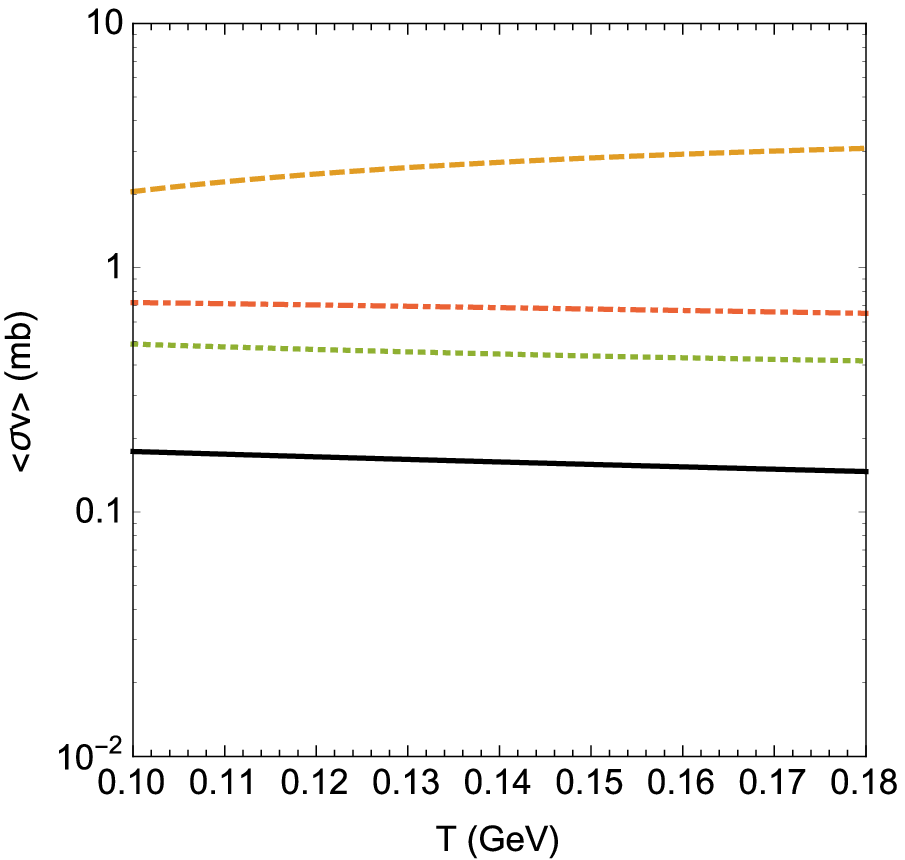}  
\hspace{0.6cm}  
\includegraphics[width=7.0cm]{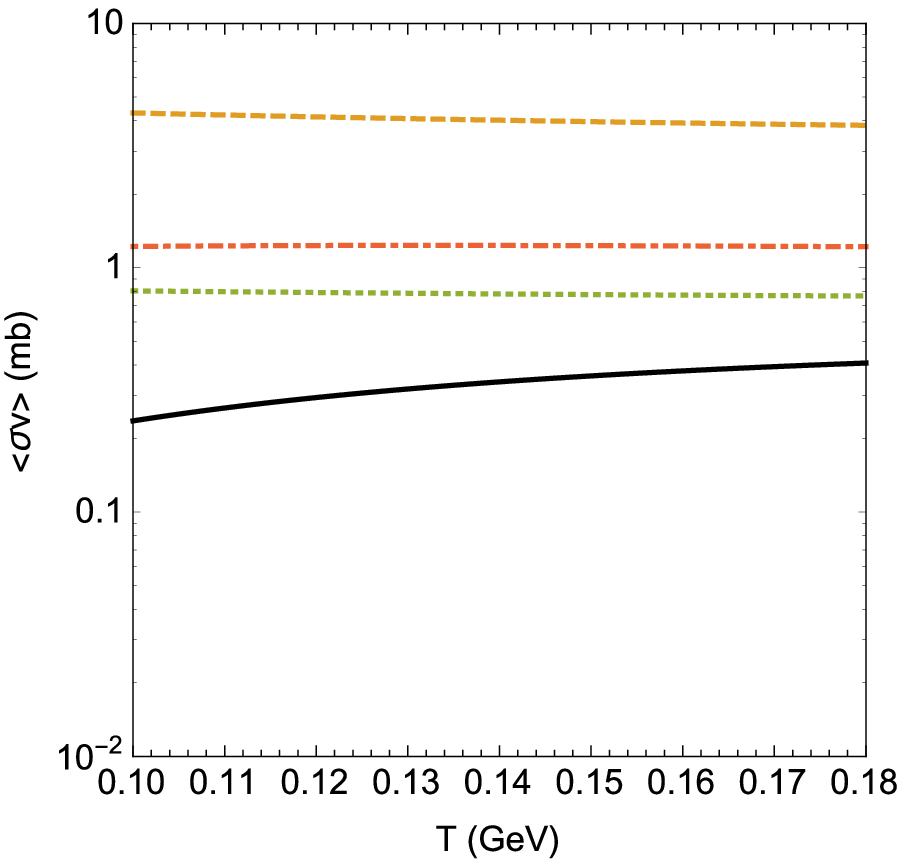}                    
\caption{$J/\Psi$ absorption and production cross sections  
by $K$'s and $K^{\ast}$'s as a  function  of the temperature.  
Top-left panel:  absorption reactions with $K  J/ \Psi$ in the initial state.   
$K J/ \Psi  \rightarrow  D_s         \bar{D}         $ (solid line),  
$K J/ \Psi  \rightarrow  D_s^{\ast}  \bar{D}^{\ast}  $ (dashed line),  
$K J/ \Psi  \rightarrow  D_s^{\ast}  \bar{D}         $ (dotted line) and  
$K J/ \Psi  \rightarrow  D_s         \bar{D}^{\ast}  $  (dot-dashed line).
Top-right panel:  production reactions with $K J/ \Psi $ in the final state. 
The line convention is the same as in the left panel. 
Bottom-left panel: absorption reactions with $K^{\ast} J/ \Psi$ in the initial state.   
$K^{\ast} J/ \Psi  \rightarrow  D_s         \bar{D}        $ (solid line),  
$K^{\ast} J/ \Psi  \rightarrow  D_s^{\ast}  \bar{D}^{\ast} $ (dashed line),  
$K^{\ast} J/ \Psi  \rightarrow  D_s^{\ast}  \bar{D}        $ (dotted line) and  
$K^{\ast} J/ \Psi  \rightarrow  D_s         \bar{D}^{\ast} $ (dot-dashed line). 
Bottom-right panel: production reactions with $K^{\ast} J/ \Psi$ in the final state.   
The line convention is the same as in the left panel. 
}
\label{thermal-strange}  
\end{figure}  
 \end{widetext}

In the upper panels of Fig.~\ref{thermal-strange}   we plot the thermally 
averaged cross sections for the $K J/ \Psi $ absorption (on the left) and 
production (on the right). As before, the  production reactions have larger cross 
sections than the corresponding inverse reactions. 
Finally, the thermally averaged cross sections for the $K^{\ast} J/ \Psi $ 
absorption and production are plotted in the lower panels of  
Fig.~\ref{thermal-strange}. It can be seen that the $J/\Psi$ 
production cross sections 
are always larger than the respective absorption cross sections. 

\section{Time evolution of the $J/\Psi$ abundance }
\label{TimeEv}
We complete this study by addressing the time evolution of the $J / \psi$ abundance 
in hadronic matter, using the thermally averaged cross sections estimated in the 
previous  section.  We shall make use of the evolution equation for the  abundances 
of particles included in processes  discussed above. The momentum-integrated 
evolution equation has the form~
\cite{EXHIC,Cho:2017dcy,Cho:2010db,Cho:2015qca,Abreu:2016qci}: 

\begin{widetext}
\ben
\frac{d N_{J/\Psi} (\tau)}{d \tau} & = &  \sum_{\varphi = \pi, \rho K , K ^{\ast} } 
\left[ \langle \sigma_{   D_{(s)}  \bar{D} \rightarrow \varphi J/\Psi } v_{ D_{(s)}  
\bar{D} } \rangle n_{D_{(s)}} (\tau) N_{\bar{D}}(\tau)- \langle 
\sigma_{\varphi J/\Psi \rightarrow  D_{(s)}  \bar{D} } v_{ \varphi J/\Psi} 
\rangle n_{\varphi} (\tau) N_{J/\Psi}(\tau)  \right. \nonumber \\ 
& &  \left. + \langle \sigma_{  D_{(s)} ^{\ast} \bar{D} ^{\ast}  \rightarrow 
\varphi J/\Psi } v_{ D_{(s)} ^{\ast} \bar{D} ^{\ast}} 
\rangle n_{D_{(s)} ^{\ast}} (\tau) N_{\bar{D} ^{\ast}}(\tau)
- \langle \sigma_{\varphi J/\Psi \rightarrow  D_{(s)} ^{\ast}      
\bar{D} ^{\ast}  } v_{ \varphi J/\Psi} \rangle n_{\varphi} (\tau)  
N_{J/\Psi}(\tau) \right. \nonumber \\  & &  \left. + \langle 
\sigma_{  D_{(s)} ^{\ast} \bar{D}   \rightarrow \varphi J/\Psi } 
v_{ D_{(s)} ^{\ast} \bar{D} } \rangle n_{D_{(s)} ^{\ast}} (\tau) 
N_{\bar{D} }(\tau) - \langle \sigma_{\varphi J/\Psi \rightarrow           
D_{(s)} ^{\ast} \bar{D}  } v_{ \varphi J/\Psi} \rangle n_{\varphi} (\tau) 
N_{J/\Psi}(\tau) \right. \nonumber \\ & &  \left. + \langle \sigma_{   
D_{(s)}  \bar{D} ^{\ast}  \rightarrow \varphi J/\Psi } v_{ D_{(s)}  
\bar{D} ^{\ast} } \rangle n_{D_{(s)} } (\tau) N_{\bar{D} ^{\ast}}(\tau) - 
\langle \sigma_{\varphi J/\Psi \rightarrow  D_{(s)}  \bar{D} ^{\ast}  } 
v_{ \varphi J/\Psi} \rangle n_{\varphi} (\tau) N_{J/\Psi}(\tau)   \right] 
\nonumber \\ & & + \sum_{\varphi = \bar{\pi}, \bar{\rho}, \bar{K},  
\bar{K} ^{\ast} } \left[ \langle \sigma_{ \bar{D}_{(s)}  D \rightarrow 
\varphi J/\Psi } v_{ \bar{D}_{(s)}  D } \rangle n_{\bar{D}_{(s)}} (\tau)   
N_{D}(\tau) - \langle \sigma_{\varphi J/\Psi \rightarrow  \bar{D}_{(s)}  D } 
v_{ \varphi J/\Psi} \rangle n_{\varphi} (\tau) N_{J/\Psi}(\tau)  \right. 
\nonumber \\  & &  \left. + \langle \sigma_{  \bar{D}_{(s)} ^{\ast} D ^{\ast} 
  \rightarrow \varphi J/\Psi } v_{ \bar{D}_{(s)} ^{\ast} D ^{\ast}} \rangle 
n_{\bar{D}_{(s)} ^{\ast}} (\tau) N_{D ^{\ast}}(\tau) - \langle 
\sigma_{\varphi J/\Psi \rightarrow  \bar{D}_{(s)} ^{\ast} D ^{\ast}  } 
v_{ \varphi J/\Psi} \rangle n_{\varphi} (\tau) N_{J/\Psi}(\tau)  \right. 
\nonumber \\  & &  \left. + \langle \sigma_{  \bar{D}_{(s)} ^{\ast} D   
\rightarrow \varphi J/\Psi } v_{ \bar{D}_{(s)} ^{\ast}  D } \rangle 
n_{\bar{D}_{(s)} ^{\ast}} (\tau) N_{ D }(\tau) - \langle 
\sigma_{ \varphi J/\Psi \rightarrow \bar{D}_{(s)} ^{\ast} D  } 
v_{ \varphi J/\Psi} \rangle n_{\varphi} (\tau) N_{ J/\Psi }(\tau) \right. 
\nonumber \\ & &  \left. + \langle \sigma_{  \bar{D}_{(s)}  D ^{\ast}  
\rightarrow \varphi J/\Psi } v_{ \bar{D}_{(s)}  D ^{\ast} } \rangle 
n_{\bar{D}_{(s)} } (\tau) N_{D ^{\ast}}(\tau) - \langle \sigma_{\varphi 
J/\Psi \rightarrow  \bar{D}_{(s)}  D ^{\ast}  } v_{ \varphi J/\Psi} \rangle \right],
\label{rateeq}
\een
\end{widetext}
where $n_{\varphi} (\tau)$ are $N_{\varphi}(\tau)$ denote  the density and 
the abundances of  $\pi,\rho, K, K^{\ast}$, charmed mesons and their 
antiparticles in hadronic matter at  proper time $\tau$. From 
Eq.~(\ref{rateeq}) we observe that 
the $J/\Psi$ abundance at a proper time $\tau$ depends on the $\varphi J/\Psi$ 
dissociation rate as well as on the $\varphi J/\Psi$  production rate. We 
remark that in the rate equation we have also considered the processes 
involving the respective antiparticles, i.e. $ \bar{\varphi}   J/\Psi 
\rightarrow  \bar{D}_{(s)}  ^{(\ast)} D  ^{(\ast)} $ and $ 
\bar{D}_{(s)}^{(\ast)} D  ^{(\ast)}  \rightarrow \bar{\varphi}  J/\Psi $. 
However, these reactions have the same cross sections as the corresponding 
conjugate processes and the results reported above will be used to evaluate 
these contributions. 
We are interested in following the time evolution of the $J/\Psi$  abundance in  
the hot hadron gas produced in heavy ion collisions.  In particular, we focus 
on central Au-Au 
collisions at $\sqrt{s_{NN}} = 200$ GeV and discuss the yields. We consider 
that $\pi, \rho , K, K^{\ast}, D$ and $D^{\ast} $  
are in equilibrium. Therefore the density $n_{i} (\tau)$ can be written  
as~\cite{EXHIC,Cho:2017dcy,Cho:2010db,Cho:2015qca,Abreu:2016qci}
\ben n_{i} (\tau) &  \approx & \frac{1}{2 \pi^2}\gamma_{i} g_{i} m_{i}^2 
T(\tau)K_{2}\left(\frac{m_{i} }{T(\tau)}\right), 
\label{densities}
\een
where $\gamma _i$ and $g_i$ are respectively the fugacity factor and the  
degeneracy factor of the relevant particle. The abundance $N_i (\tau)$ is 
obtained by multiplying  the density $n_i(\tau)$ by the volume $V(\tau)$. 
The time dependence is introduced through the temperature $T(\tau)$ and 
volume $V(\tau)$ profiles appropriate to model the dynamics of relativistic 
heavy ion collisions after the end of the quark-gluon plasma phase. The 
hydrodynamical expansion and cooling of the hadron gas is modeled as  in 
Refs.~\cite{EXHIC,Cho:2017dcy,Cho:2010db,Cho:2015qca,Abreu:2016qci} by a  
the boost invariant Bjorken flow  with an accelerated transverse expansion:
\ben
T(\tau) & = & T_C - \left( T_H - T_F \right) \left( \frac{\tau - \tau _H }{\tau _F -  
\tau _H}\right)^{\frac{4}{5}} , \nonumber \\V(\tau) & = & \pi \left[ R_C + v_C 
\left(\tau - \tau_C\right) + \frac{a_C}{2} \left(\tau - \tau_C\right)^2 \right]^2 \tau_C.
\label{TempVol}
\een
In the equation above, $R_C = 8.0$ fm and $\tau _C = 5.0$ fm/c denote the final 
transverse  and longitudinal sizes of the quark-gluon plasma, while $v_C = 0.4c$ 
and  $a_C = 0.02 \, c^2$/fm are its transverse flow velocity and transverse 
acceleration at this time. 
$T_C = 175$ MeV is the critical temperature for the quark-gluon plasma to hadronic 
matter transition; $T_H = T_C = 175$ MeV is the temperature of the hadronic matter 
at the 
end of the mixed phase, occurring at the time $\tau_H = 7.5$ fm/c. The freeze-out 
temperature  $T_F = 125$ MeV then leads to a freeze-out time $\tau _F = 17.3$ fm/c.
In addition, we assume that the total number of charmed quarks in charmed hadrons is  
conserved during the processes, and that the total number of charm quarks produced 
from  the initial stage of collisions at RHIC is 3, yielding the charm quark 
fugacity factor $\gamma _C \approx 6.4$ in Eq.~(\ref{densities}) 
\cite{EXHIC,Cho:2017dcy,Cho:2010db,Cho:2015qca,Abreu:2016qci}. 
For pions and $\rho$ mesons, we follow Refs.~\cite{ChenPRC,Abreu:2016qci,ChoLee1} and 
assume that their total number at freeze-out is 926 and 68, respectively. 
In the case of $K^{(\ast)}$ and $\bar{K}^{(\ast)}$ mesons \cite{EXHIC}, we work with the 
assumption that strangeness reaches approximate chemical equilibrium in heavy ion    
collisions due to the short equilibration time in the quark-gluon plasma and the net 
strangeness of the QGP is zero.

In the context of the statistical model, hadrons are in thermal and chemical 
equilibrium when they are produced at chemical freeze-out in heavy ion collisions. 
Thus, the  $J/\Psi$ yield at the end of the mixed phase is
\ben 
N_{J/\Psi} & \approx &  \frac{1}{2 \pi^2} \gamma_{c} ^2  
g_{J/\Psi} m_{J/\Psi}^2 T_H K_{2} \left(\frac{m_{J/\Psi} }{T_H}\right) 
V(\tau_H) \nonumber \\ 
&\approx &  1.722 \times 10^{-2} . 
\label{NJPsi}
\een

\begin{figure}[th]
\centering
\includegraphics[width=8.0cm]{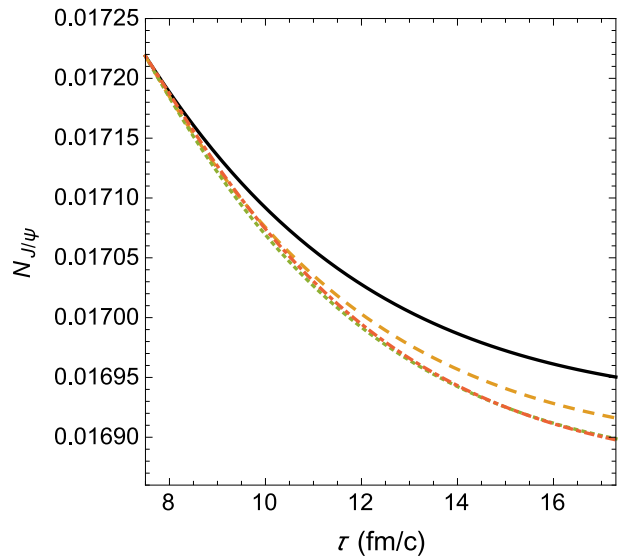}
\caption{ Time evolution of $J/\Psi$ abundance as a function of the proper  
time in  central Au-Au collisions at $\sqrt{s_{NN}} = 200$ GeV. Solid, dashed, 
dotted, dot-dashed  lines represent the situations with only $ \pi - J/\Psi$ 
interactions and also adding the  $\rho - J/\Psi$, $ K - J/\Psi$ and 
$ K^{\ast} - J/\Psi$ contributions, respectively.}
\label{TimeEvolJPsi}
\end{figure}

In Fig.~\ref{TimeEvolJPsi} we present the time evolution of the $J/\Psi$ 
abundance as a function of the proper time in central Au-Au collisions at 
$\sqrt{s_{NN}} = 200$ GeV.  Looking at the evolution equation, Eq. (\ref{rateeq}),
we can see that the fate of the $J/\Psi$ population will be determined by the 
production and absorption cross sections and by the multiplicities of the other 
mesons, especially the pion multiplicity. While the cross sections alone would favor 
an enhancement of the $J/\Psi$ yield, the relative multiplicities favor its 
reduction, since in the hadron gas there are much more pions and kaons 
(which hit and destroy the charmonium states) than $D$'s, $\bar{D}$'s, $D_s$'s
and $\bar{D_s}$'s (which can collide and create them). The result of this competition 
is an approximate equilibrium between production and absorption. The  $J/\psi$ yield 
remains nearly constant during the hadron gas phase.  From the solid line in the  
figure we can see that if there were only pions in the gas, there would be a small 
suppression of the $J/\Psi$ yield. This comes from a cancellation between a large 
difference in the cross sections (the upper panels in Fig.~\ref{thermal-light}) 
favoring production with a large difference in multiplicities, as pions are much more 
abundant than open charm mesons. 
Approximately the same cancellation occurs if the gas 
would include $\rho$'s, kaons and $K^*$'s.

In view of the uncertainties inherent to our calculations, all these numbers 
contain errors and should not be taken as definitive. 
A short list of the sources of uncertainties would certainly include the following 
items: 

\vspace{0.3cm}
\noindent
i) The use of 
the SU(4) Lagrangian, which governs the interactions between  particles. It could be 
replaced by some other theory. This would change the absolute values of the matrix 
elements of the reactions considered. We are primarily interested in the 
equilibruim (or absence of) between the absorption reactions and the corresponding 
productions reactions. A simple change in the magnitude of the matrix elements would 
not affect the final equilibrium, since they would be still connected by the 
same detailed balance relations. Increasing production will increase absorption 
in the same proportion. 

\vspace{0.3cm} 
\noindent 
ii) The form factors. Both their functional form and 
the cutoff values could be changed. In fact a small change in the cutoff parameters 
would already transform the lines in Fig.~\ref{TimeEvolJPsi} into bands. However, 
as far as net changes in the $J/\Psi$ multiplicity are concerned, the same discussion
of  item i) applies here.

\vspace{0.3cm} 
\noindent 
iii) The parametrization of the hydrodynamical expansion, Eqs.(\ref{TempVol}), could 
be changed by a more realistic one. This could make the system cool faster or 
slower and consequently change the multiplicities of the different particles 
($\pi$'s, $D$'s, ...etc.) in different ways. This could potentially reverse the 
direction of the dynamics. For example, increasing the number of pions  with respect 
to the number of  open charm mesons would increase the absorption of $J/\Psi$'s. 

In view of the above discussion, we conclude that, even though there are still many
aspects to be considered and/or improved, we believe that our main result, the 
approximate constancy of the number of $J/\Psi$'s throughout the hadron gas phase, 
is not likely to be dramatically changed.  If confirmed, this result is very 
interesting for the physics of the quark gluon plasma, since  $J/\Psi$ production 
will be entirely determined by the QGP dynamics.

\section{Concluding Remarks}
\label{Conclusions}

Precise measurements of $J/\psi$ abundancies in heavy ion collisions are 
an important source of information about the properties of the quark-gluon          
plasma phase. During this phase $J/\psi$ is produced by recombination 
of charm-anticharm pairs. However, after  hadronization
the $J/\psi$'s  interact with other hadrons in the expanding  hadronic matter. 
Therefore, the $J/\psi$'s can be  destroyed  in collisions with  other comoving  
mesons, but they can also be produced through the inverse reactions.
In order to evaluate the hadronic effects on the $J/\psi$ abundance in 
heavy ion collisions one needs to know the $ J/\psi$ cross sections with 
other mesons. 

In this work we have studied $J/\psi$ dissociation and production reactions, making 
use of effective field Lagrangians to obtain the cross sections for the processes  
$(\pi , \rho, K,  K ^{\ast})  + J/ \psi \rightarrow D_{(s)}  \bar{D}  ,  
D_{(s)} ^{(\ast)} \bar{D} ,  D_{(s)}  \bar{D} ^{(\ast)} , D_{(s)} ^{\ast} 
\bar{D} ^{\ast} $ 
and the corresponding inverse processes. We have then   
computed the  thermally averaged cross sections for the dissociation and       
production reactions, the latter being larger.  Finally, we have used the 
thermally averaged cross sections as input in a rate equation and have followed
the evolution of the $J/\psi$ abundance in a hadron gas. 

With respect to the existing calculations, the improvements introduced here 
are the inclusion of $K$ and $K^*$'s in the effective Lagrangian approach 
(and the computation of the corresponding cross sections) 
and the inclusion of processes involving the new exotic charmonium states 
$Z_c(3900)$ and $Z_c(4025)$. 

We conclude that the interactions between $J/\Psi$ and  all the 
considered  mesons do not significantly change the original $J/\Psi$ abundance, 
determined at the end of the quark gluon plasma phase. Consequently, any 
significant change in the $J/\psi$  abundance comes from  dissociation and  
regeneration processes in the QGP phase. 

\begin{acknowledgements}
We are deeply grateful to R. Rapp for reading our manuscript and making several 
enlightening observations.  
The authors would like to thank the Brazilian funding agencies CNPq (contracts
310759/2016-1 and  311524/2016-8)
FAPESP (contracts 12/50984-4 and 17/07278-5) for financial support.
\end{acknowledgements} 

\begin{thebibliography}{99}
\bibitem{correc}           We have identified two misprints in the expressions  
                           of  some  amplitudes $\mathcal{M}^{(K^{\ast})}$ which 
                           appear in Ref.~\cite{reginas}: in the amplitude 
                           $M_{2b}$, third line in Eq. (22), the sign of the 
                           term $2 p_4 ^\mu$ should be a minus sign; in the 
                           amplitude $M_{4d}$, sixth line in Eq. (22), the 
                           $p_{4 \rho}$ should be $p_{3 \rho}$.
\end{thebibliography}
%
%

\end{document}